\documentclass[10pt,aps,prd,twocolumn,showpacs,superscriptaddress,nofootinbib,nobibnotes,longbibliography,floatfix]{revtex4-1}

\usepackage{amsmath}
\usepackage{graphicx} 
\usepackage[utf8]{inputenc}
\usepackage[T1]{fontenc}
\usepackage{enumitem}
\usepackage{bm}
\usepackage{mathtools,amsmath,amssymb,amsfonts,graphicx,tensor,csquotes,accents, commath,chngcntr,siunitx}
\usepackage[dvipsnames]{xcolor}
\usepackage{hyperref}
\hypersetup{colorlinks=true, citecolor=MidnightBlue,
            linkcolor=MidnightBlue, urlcolor=MidnightBlue, linktocpage=true}
\usepackage[normalem]{ulem}
\usepackage{tensor}
\usepackage{leftindex}
\usepackage{caption}
\usepackage{subcaption}
\usepackage{amssymb} 
\usepackage{orcidlink}
\usepackage{pifont}
\usepackage{lmodern}
\usepackage{cancel}

\usepackage{float}

\usepackage{array}

\usepackage{xcolor}

\usepackage{tabularray}
\UseTblrLibrary{booktabs}

\newcommand{\sissa}{SISSA, Via Bonomea 265, 34136 Trieste, Italy \&  INFN Sezione di Trieste}

\newcommand{\ifpu}{IFPU - Institute for Fundamental Physics of the Universe, Via Beirut 2, 34014 Trieste, Italy}

\newcommand{\centra}{CENTRA, Departamento de Física, Instituto Superior Técnico – IST,
Universidade de Lisboa – UL, Avenida Rovisco Pais 1, 1049-001 Lisboa, Portugal}

\begin{document}

\title{Systematic bias in LISA ringdown analysis due to waveform inaccuracy }

\author{Lodovico Capuano~\orcidlink{0009-0001-0369-6635}}
\email{lcapuano@sissa.it}
\affiliation{\sissa}
\affiliation{\ifpu}

\author{Massimo Vaglio~\orcidlink{0000-0002-7285-3489}}
\email{mvaglio@sissa.it}
\affiliation{\sissa}
\affiliation{\ifpu}

\author{Rohit S. Chandramouli~\orcidlink{0000-0001-5229-2752}}
\email{rchandra@sissa.it}
\affiliation{\sissa}
\affiliation{\ifpu}

\author{Chantal Pitte~\orcidlink{0009-0009-0524-7292}}
\email{cpitte@sissa.it}
\affiliation{\sissa}
\affiliation{\ifpu}

\author{Adrien Kuntz~\orcidlink{0000-0002-4803-2998}}
\email{adrien.kuntz@tecnico.ulisboa.pt}
\affiliation{\centra}

\author{Enrico Barausse~\orcidlink{0000-0001-6499-6263}}
\email{barausse@sissa.it}
\affiliation{\sissa}
\affiliation{\ifpu}

\begin{abstract}

Inaccurate modeling of gravitational-wave signals can introduce systematic biases in the inferred source parameters. 
As detector sensitivities improve and signals become louder, mitigating such waveform-induced systematics becomes increasingly important. 
In this work, we assess the systematic biases introduced by an incomplete description of the ringdown signal from massive black hole binaries in the LISA band. Specifically, we investigate the impact of mode truncation in the ringdown template. Using a reference waveform composed of 13 modes, we establish a mode hierarchy and determine the minimum number of modes required to avoid parameter biases across a wide range of LISA sources. For typical systems with masses $\sim 10^6$--$10^7\,M_\odot$ at redshifts $z \sim 2$--$6$, we find that at least 3--6 modes are needed for accurate parameter estimation, while high-SNR events may need at least 10 modes. Our results are a window-insensitive lower bound on the minimum number of modes, as more modes may be needed depending on the choice of time-domain windowing of the post-merger signal. 
\end{abstract}

\maketitle

\section{Introduction}
As a binary black hole (BH) system emits gravitational waves (GWs), the two black holes spiral closer together, ultimately merging in a coalescence event. 
Immediately after the merger, the remnant BH is in a perturbed state and continues to emit GWs in a ringdown phase, gradually settling into a stable configuration. 
The GW signal in this regime is well understood in terms of BH perturbation theory. 
In general, any perturbation of the spacetime metric of a rotating BH in General Relativity (GR), at linear order, is governed by the Teukolsky equation~\cite{Teukolsky:1973ha}.
Solutions to the Teukolsky equation, given appropriate boundary conditions, are expressed as a 
linear superposition of quasi-normal modes (QNMs), i.e. exponentially damped sinusoids~\cite{1970Natur.227..936V,89c00a43-362f-3c19-9cc2-fa0808f91935, Anninos_1993, Kokkotas_1999, Berti_2009, Konoplya_2011}.
Specifically, QNMs are characterized by complex frequencies and  amplitudes, with the imaginary part of the frequency determining the damping timescale.

Linear superpositions of QNMs provide a valid description of the ringdown at intermediate times after the merger.
At later times, the GW signal is instead dominated by a non-oscillatory power-law tail~\cite{DeWitt:1960fc, Ma:2024hzq, DeAmicis:2024eoy}, while at earlier times close to the merger,
non-linear effects determine the signal behavior~\cite{Bhagwat:2017tkm,Okounkova:2020vwu}.
The QNM-dominated stage of the ringdown is also affected by (small) contributions beyond linear perturbation theory, such as quadratic QNMs~\cite{London_2014, Mitman:2022qdl, Cheung:2022rbm,Bucciotti:2023ets, Lagos_22, Bucciotti:2024jrv, Bucciotti:2024zyp, Zhu:2024rej, Redondo-Yuste:2023seq, Yi:2024elj, Bucciotti:2025rxa, Ma:2024qcv, Khera:2024yrk}. Moreover, third-order effects in perturbation theory, such as the dynamical evolution of the BH mass and the ensuing modification of the spectrum, have also been discussed in the recent literature~\cite{PhysRevD.105.064046,PhysRevD.109.044048,PhysRevD.110.084081}.

Observation of the BH ringdown allows the extraction of crucial physical information,  related to the astrophysical properties of the source and to fundamental physics~\cite{Babak:2010ej,Sesana:2021jfh,LISA:2022kgy}.
With improvements in ground-based GW detectors and upcoming space-based detectors, we are likely to be able to detect many GW events with a high signal-to-noise ratio (SNR)~\cite{KAGRA:2013rdx,Reitze:2019iox,Punturo:2010zz,LISA:2017pwj,LISA:2024hlh,Luo:2021qji,Hu:2024vvu,TianQin:2015yph}.
In particular, the Laser Interferometer Space Antenna (LISA)~\cite{LISA:2017pwj,LISA:2024hlh}
is expected to observe anywhere from a few to thousands of massive BH binaries per year (depending on the astrophysical model)~\cite{Klein:2015hvg, Bonetti:2018tpf, Barausse:2020mdt, 2019MNRAS.486.2336D,Barausse:2023yrx, Izquierdo-Villalba:2024bhc}. Many of these events could have SNRs $\gtrsim 100$ in the ringdown phase~\cite{Berti:2016lat,Yi:2024elj}, enabling precise parameter estimation with statistical uncertainties far smaller than those achievable with the current LIGO-Virgo-KAGRA network. 
The reduction in statistical errors can potentially expose systematic biases induced by waveform mismodeling of the GW signal.
Such mismodeling can arise because of unaccounted effects from the astrophysical environment or new physics~\cite{Barausse:2014tra}, but also from inaccurate or incomplete modeling of known dynamics or truncation in perturbation methods~\cite{Yunes:2009ke, Cutler:2007mi, Vallisneri:2012qq, Sampson:2013jpa, Favata:2013rwa, Saini:2022igm, Read:2023hkv, Owen:2023mid, Pitte:2023ltw, Kejriwal:2023djc, Gupta:2024gun, Chandramouli:2024vhw, Garg:2024qxq, Yi:2025pxe, Gupta:2024gun}.


In this work, we focus on the accuracy of ringdown waveforms, which is needed to perform BH spectroscopy~\cite{Kokkotas_1999, Dreyer:2003bv, Berti_2006} and tests of GR~\cite{PhysRevD.40.3194, Berti:2018vdi, Berti_2009, Toubiana:2023cwr}.
Specifically, we analyze the systematic biases induced by capping the number of modes in the ringdown waveform template.
Indeed, given an observed ringdown GW signal in LISA, it is not obvious \textit{a priori} how many modes one should include in the  template: too many modes can lead to overfitting issues, while including too few will bias parameter estimation. 
Starting with a template containing a maximal set of modes $N_{\text{max}}$, which we take as our fiducial (or ``true'') waveform, we compare it to approximate templates, including only the first $N < N_{\text{max}}$ highest-SNR modes.  
As will be discussed in Sec.~\ref{sec2}, we use fits of numerical relativity (NR) simulations for 11 linear modes and 2 quadratic modes, for a total of $N_{\text{max}} = 13$ modes. 
We estimate systematic errors using the linear signal approximation~\cite{Flanagan:1997kp,Cutler:2007mi,Lindblom:2008cm},  comparing them with the statistical errors obtained from the Fisher approximation~\cite{Fisher:1922saa}.
We determine the minimum number of modes $N_{\min}$ required for the estimation of unbiased parameters and explore how this threshold varies for different systems.
We compute $N_{\min}$ in two equivalent ways, by checking if the following criteria are satisfied: (i) systematic errors are smaller than the statistical errors, (ii) the mismatch is below an SNR-dependent threshold.
Since we work in the frequency domain, we introduce a high-frequency cutoff to avoid windowing artifacts, such as spectral leakage and mode contamination.
Thus, our $N_{\min}$ is a window-insensitive lower bound (underestimate) of the number of modes needed for accurate parameter estimation.

 
The work is organized as follows. In Sec.~\ref{sec3} we describe the details of our ringdown model. In Sec.~\ref{sec2} we provide an overview of the formalism used to assess waveform inaccuracy. 
Finally, in Sec.~\ref{sec4} we present and discuss our results. The appendices contain robustness checks of our calculations and further dependencies of the results on the binary parameters. In Appendix~\ref{Linear_regime} we explicitly check the validity of the linear signal approximation introduced in Sec.~\ref{sec3}. In Appendix~\ref{NR_comparison} we discuss the issue of the starting time. In Appendix~\ref{app_spins} we study the dependence of $N_\text{min}$ on the individual spins, mass ratio, inclination and angular position of the source. In Appendix~\ref{appD}, we compare our findings, obtained with a single high-frequency cutoff for all modes, with a different approach, in which a phenomenological mode-dependent tapering in the frequency domain is introduced. We also validate the frequency domain predictions by comparing the results with that for a time domain model.
We work in geometric units with $c = G = 1$.
We also use Latin indices for vector/tensor components and boldface symbols for abstract vectors.

\section{The ringdown model}
\label{sec3}

\subsection{Time and frequency-domain models}

The ringdown signal in the time domain can be modeled as an infinite sum of damped sinusoids. Specifically, we write the GW strain in time domain for $t>t_0$ (where $t_0$ is the starting time of the ringdown) as
\begin{align} \label{eq:QNMModelTimeDomain}
    h_+ &- i h_\times = \frac{M}{d_L} \sum_{\ell mn} \bigg[ \mathcal A_{\ell mn} e^{-i \omega_{\ell mn} (t-t_0)} e^{- (t-t_0)/\tau_{\ell mn}} \nonumber \\
    &+ \mathcal A_{\ell mn}' e^{i \omega_{\ell -mn}  (t-t_0)} e^{- (t-t_0)/\tau_{\ell -mn}} \bigg] \vphantom{a}_{-2}Y_{\ell m}(\iota, \varphi)
\end{align}
where $M$ is the remnant BH mass in the detector frame, $d_L$ is the luminosity distance to the source,  $\omega_{\ell mn}$ and  $-1/\tau_{\ell mn}$ are the real and imaginary parts of the QNM frequency, and $(\iota, \varphi)$ are the angular coordinates of the direction of propagation as seen from the remnant BH, with $\prescript{}{-2}{Y_{\ell m}}$ the   spin-weighted spherical harmonics.
Note that for a given set of $(\ell mn)$, there are two mode contributions $\mathcal A_{\ell mn}$ and $\mathcal A'_{\ell mn}$: this is because 
for any given QNM with frequency $\omega_{\ell mn}$ and damping time $\tau_{\ell mn}$, there exists a ``mirror mode'' also solving the equations of motion, which corresponds to the second term in Eq.~\eqref{eq:QNMModelTimeDomain}~\cite{Berti_2009, Berti:2005ys}. Finally, for this study, we set the starting time to $t_0=20M$ after the luminosity peak. For further details on the choice of the starting time, see App.~\ref{NR_comparison}

We will further assume, as  common in the literature~\cite{Berti:2007fi, Berti:2007zu}, that the amplitude of the mirror modes is given by 
\begin{equation} \label{eq:mirrorModeAmplitude}
    \mathcal A'_{\ell mn} = (-1)^\ell \mathcal A_{\ell -mn}^* ,
\end{equation}
where $*$ denotes complex conjugation.
This property follows from equatorial symmetry~\cite{Bucciotti:2024jrv,Isi:2021iql}, i.e. from neglecting orbital precession in the merger waveform.\footnote{This 
may  be only approximately true for precessing sources, but a full characterization of the amplitude of the mirror modes for precessing systems has not yet been considered in the literature.}

In the following, we define the Fourier transform of a time series\footnote{When the time domain data is real, the Fourier components satisfy the relation $\tilde{u}(f)=\tilde{u}^*(-f)$.} $u(t)$ as
\begin{equation}
    \tilde{u}(f)=\int_{-\infty}^{+\infty}{\rm d}t\,u(t)e^{2\pi i f t}\,.
\end{equation}

Furthermore, to compute the Fourier transform, following~\cite{Flanagan:1997kp,Berti:2005ys},
we continuously extend the ringdown for $t<t_0$ by symmetrically mirroring the waveform across $t=t_0$. More precisely, we replace $e^{-(t-t_0)/\tau} \rightarrow e^{(t-t_0)/\tau}$ for $t<t_0$, and divide the amplitude by $\sqrt{2}$ to avoid double-counting. By ensuring continuity of the signal in $t=t_0$, this approach improves the behavior of the Fourier transform at high frequencies, preventing the introduction of spurious Fourier components. This procedure gives: 
%
%
\begin{align} \label{eq:ringdownModelFreqDomain}
    \tilde h_{+,\times}(f) &= e^{i \omega t_0} i^{\frac{1\mp1}{2}}\frac{M}{\sqrt{2}d_L} \nonumber \\
    &\times \sum_{\ell mn} \frac{A_{\ell mn} Y_{\ell m}^{+,\times}}{\tau_{\ell mn}} \left( L_{\ell m n}^+ \pm L_{\ell m n}^-  \right)\,,
\end{align}
where we separate the complex amplitudes in a modulus and a phase, $\mathcal{A}_{\ell m n} = A_{\ell m n} e^{i \Phi_{\ell m n}}$, and we defined the angular functions
\begin{equation}
     Y_{\ell m}^{+,\times} = \prescript{}{-2}{Y_{\ell m}}(\iota, 0) \pm (-1)^\ell \prescript{}{-2}{Y_{\ell-m}}(\iota, 0)\,.
\end{equation}
The Lorentzian functions read
\begin{equation}
    L_{\ell m n}^{\pm} =\frac{e^{\pm i(m\varphi + \Phi_{\ell mn})}}{\tau_{\ell mn}^{-2} + (2\pi f \mp \omega_{\ell mn})^2} \,.
\end{equation}
{Assuming that the detector response is time-independent, and given two times series $u(t)$ and $v(t)$, we define the matched filter inner product as
\begin{equation}
    \left(u\left|\right. v\right)\equiv 4 \mathrm{Re}\left[\int_0^\infty{\rm d}f\,\frac{\tilde{u}(f)\tilde{v}^*(f)}{S_n(f)}\right]\,.
    \label{inner_product}
\end{equation}
In Eq.~\eqref{inner_product}, $S_n(f)$ is the (one-sided) noise power-spectral density (PSD) of the detector given in~\cite{LISA:2017pwj}. 

The signal measured by the detector also depends on the response of LISA to the incoming GW, which is encoded in the time-dependent LISA transfer function~\cite{Cornish:2002rt,Marsat:2018oam}. For typical sources observed by LISA, the ringdown lasts from minutes to hours, which is negligible compared to the timescale of LISA's motion, thereby making the stationary approximation sufficient for our work.
In the stationary approximation, the LISA configuration is effectively described by two detectors~\cite{Cutler:1997ta,Cornish:2002rt}, which we will label by $\mathrm{I,II}$.
The LISA response is then characterized by the antenna response functions (for each GW polarization) of the two detectors, which are given by~\cite{Cutler:1997ta,Cornish:2002rt}
\begin{equation}
    F_{+,\times}^{\text{II}}\left( \theta, \phi, \psi \right)=F_{+,\times}^{\text{I}}\left( \theta, \phi-\pi/4, \psi \right)\,.
\end{equation}
The antenna response functions depend on the sky position of the source, determined by the angles $\theta$ and $\phi$, and a polarization angle $\psi$, and their expressions read~\cite{Maggiore:2007ulw}
\begin{equation}
\begin{split}
    F_+\left( \theta, \phi, \psi \right) =& \frac{\sqrt{3}}{4} \left( 1 + \cos^2 \theta \right) \cos 2\psi \cos 2\phi \,,\\
    &-\frac{\sqrt{3}}{2} \cos \theta \sin 2\psi \sin 2\phi \\
    F_\times\left( \theta, \phi, \psi \right) =& \frac{\sqrt{3}}{4} \left( 1 + \cos^2 \theta \right) \sin 2\psi \cos 2\phi \\
    &+ \frac{\sqrt{3}}{2}\cos \theta \cos 2\psi \sin 2\phi\,.
\end{split}
\end{equation}
The signal measured in a single detector is then given by
\begin{equation}
    \tilde h(f)^{\mathrm{I/II}}=F_+^{\mathrm{I/II}} \tilde h_+(f)+F_\times^{\mathrm{I/II}} \tilde h_\times(f)\,.
\end{equation}
For the rest of the paper, to avoid cluttering, we will drop the tilde indicating the Fourier transform.

\subsection{Validity regime of the QNM model}
To assess the number of modes needed to ensure waveform accuracy, it is important to account for the validity regime of the frequency-domain QNM model given by Eq.~\eqref{eq:ringdownModelFreqDomain}.
The time-domain damped-sinusoid QNM model given by Eq.~\eqref{eq:QNMModelTimeDomain} starts being valid only   a short time after the peak of the strain amplitude, because early times are dominated by the prompt response, and ceases to be valid at late times, where nonlinear tails become significant~\cite{Berti:2025hly,PhysRevD.34.384,Price_tails,Cheung:2023vki,Zhu:2023mzv,DeAmicis:2024eoy}.
In general, to isolate the regime where the QNM model is valid, one needs to apply a window function $W(t)$ to the time-domain signal that removes the early and late time contributions. 
As a result, the frequency-domain response corresponds to a convolution of Eq.~(\ref{eq:QNMModelTimeDomain}) with the Fourier transform of the window $W(f)$.

The bandwidth of $W(f)$ is roughly given by $\Delta f \sim 1/\Delta t$, with $\Delta t$ being the transition time of the window, characterizing how fast the latter rises or drops.
A very sharp transition with small $\Delta t$ would thus result in a large frequency spread of the signal response, leading to spectral leakage.
In addition, if the window is discontinuous in its $p$-th derivative, the high-frequency fall-off goes as $1/f^{p+1}$ (see Appendix~\ref{app:tapering} for details).
Thus, the smoothness of the window is crucial when trying to mitigate spectral leakage.

Since the choice of window is not unique,
in Fig.~\ref{fig:QNM-leakage} we consider a typical ringdown system and
compare  the frequency-domain waveforms 
obtained with the standard Heaviside window and with the mirroring technique adopted in our work.
 Since the Heaviside filter and the mirroring produce a discontinuity at $t=t_0$ respectively in the strain and in its first derivative, the corresponding frequency falloffs are $1/f$ and $1/f^2$.
We find that with the Heaviside window, there is significant spectral leakage from the 220 mode, which contaminates the peaks of the higher harmonics, such as the 330 and 550 modes shown in Fig.~\ref{fig:QNM-leakage}.
With the mirroring, the 330 mode is much less contaminated, due to suppressed spectral leakage from the 220. 
However, due to its smaller amplitude, the 550 is still significantly contaminated. 
More in general, depending on the choice of window function, spectral leakage can contaminate the frequency-domain description of the sub-dominant higher frequency modes.

\begin{figure}[t]
    \centering
    \includegraphics[width=\linewidth]{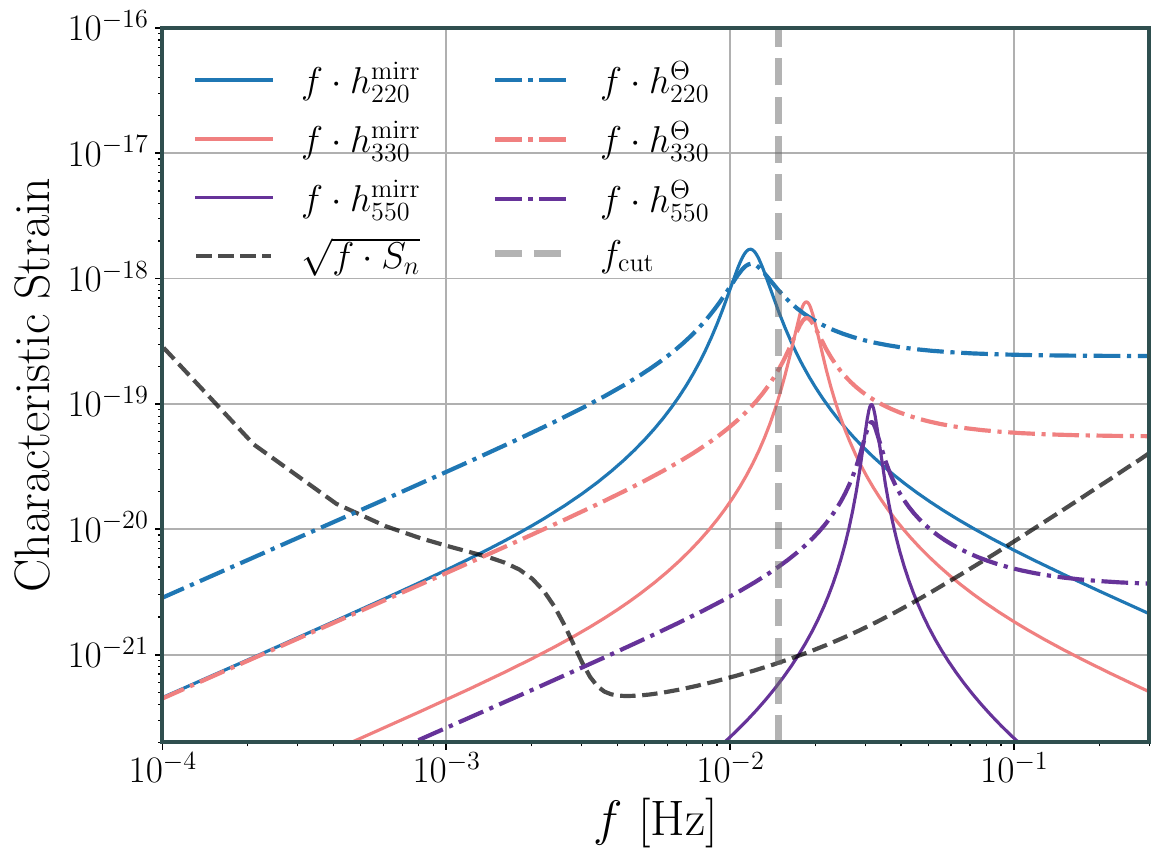}
    \caption{Characteristic strain of three QNMs in the ringdown of a BBH with remnant mass \( M = 1.44\times 10^6 M_\odot \) and dimensionless spin \( a = 0.66 \), observed at redshift \( z = 0.1 \). The modes \( h_{220} \), \( h_{330} \), and \( h_{550} \) are shown in light blue, pink, and purple, respectively. Solid and dot-dashed lines indicate the mirroring technique ($h^{\rm mirr}_{\ell m n}$) and a Heaviside time window ($h^{\Theta}_{\ell m n}$), applied before taking the Fourier transform. The black dashed line represents the LISA strain sensitivity curve, while the thick grey vertical dashed line marks indicates the PhenomA $f_\text{cut}$. Observe that there is significant spectral leakage from the dominant \( h_{220} \) mode into higher-frequency regions above $f_\text{cut}$. 
    }
    \label{fig:QNM-leakage}
\end{figure}

In practice, to deal with spectral leakage, we use a high-frequency cutoff that ensures that our model is not sensitive to the choice of window.
The price that we pay is the loss of information from some sub-dominant higher modes that are potentially contaminated by spectral leakage from the dominant ones.
We adopt the PhenomA~\cite{Babak:2006ty} high-frequency cutoff for the frequency-domain model, given by
\begin{equation}
    f_{\text{cut}}=\frac{\sum_{i=0}^2c_i\eta^i}{\pi (m_1+m_2)}\,,
    \label{eq:cutoff}
\end{equation}
where $c_i$ are numerical coefficients given in~\cite{Babak:2006ty}, and we introduced the symmetric mass ratio $\eta = m_1 m_2/(m_1+m_2)^2$.  
In Fig.~\ref{fig:QNM-leakage}, we see that for $\omega_{220}/(2\pi)<f<f_{\rm cut}$, the frequency-domain model is not sensitive to the choice of window. 
For $f>f_{\rm cut}$ on the other hand, the frequency-domain model becomes window-dependent.
This cutoff is particularly important for high-mass sources with $M \gtrsim 10^8 M_{\odot}$, as the QNM high-frequency tails will match the low-frequency behavior of the LISA sensitivity curve. 
Note that $f_{\rm cut}$ is effectively set by the 220 mode, as it dominates the spectral leakage.  
Specifically, the model is truncated  where the 220 mode begins to exhibit windowing artifacts in the frequency domain.  
While each individual mode is valid within approximately \( \sim 1/\tau_{\ell m n} \) of its peak, a common cutoff must be applied across all modes, since the model (and the data) are constructed as sums over them.


To mitigate spectral leakage in a more systematic way, one could optimally tune the transition width $\Delta t$ for a $C^{\infty}$ window (such as the Planck-taper~\cite{McKechan:2010kp}). This approach is discussed in Appendix~\ref{app:tapering}.
However, our practical approach with the frequency cutoff still allows us to obtain results that are insensitive to the choice of window. 
Thus, in our work, we essentially estimate a window-insensitive lower bound for the number of modes $N_{\min}$ needed for model accuracy.
In Appendix~\ref{app:tapering}, we show results computed from a phenomenological implementation of a tapered frequency-domain ringdown model that captures features of an optimal window function.
We find that  the results for $N_{\min}$ obtained with the frequency cutoff of the mirrored frequency-domain response are indeed generically smaller than those obtained with the phenomenological optimal taper.
This confirms that our results for $N_{\min}$ presented in the main text should be regarded as lower bounds, independent of the choice of window. We have further validated our results by comparing them to those obtained with an analysis of the mismatch in the time domain, for a specific point of the parameter space $(M,z)$ (c.f. Appendix~\ref{app:tapering}). 

\subsection{Modes}
\label{sec2c}
\begin{table*}[htbp]

  \centering
  \resizebox{0.7\textwidth}{!}{
  \begin{tblr}{
      colspec={lllllll},
      row{1}={font=\bfseries},
      column{1}={font=\itshape},
      row{even}={bg=gray!10},
      cells={c}
    }
    
              & Linear  & Quadratic  & Fundamental  & Overtone & Prograde &Retrograde  \\
    \toprule
    \toprule
    220 & \checkmark &  & \checkmark & & \checkmark &  \\
    210 & \checkmark &  & \checkmark & & \checkmark & \\
    330 & \checkmark &  & \checkmark & & \checkmark & \\
    320 & \checkmark &  & \checkmark & & \checkmark & \\
    440 & \checkmark &  & \checkmark & & \checkmark & \\
    550 & \checkmark &  & \checkmark & & \checkmark & \\
    \bottomrule
    221 & \checkmark &  &  & \checkmark & \checkmark &  \\
    211 & \checkmark &  &  & \checkmark & \checkmark & \\
    331 & \checkmark &  &  & \checkmark & \checkmark & \\
    \bottomrule
    r220 & \checkmark &  & \checkmark &  &  &  \checkmark\\
    r210 & \checkmark &  & \checkmark &  &  & \checkmark\\
    \bottomrule
    220$\times$ 220 &  & \checkmark & \checkmark &  & \checkmark &  \\
    220$\times$ 330 &  & \checkmark & \checkmark &  & \checkmark & \\
    \bottomrule
    \bottomrule
  \end{tblr}
  }
  \caption{Modes included in our model, classified according to their order in perturbation theory, their overtone number and their rotation with respect to the BH spin.}
  \label{table_modes}
\end{table*}
In our work, we use
the frequency domain waveform of Eq. \eqref{eq:ringdownModelFreqDomain} with the 13 modes listed in Table~\ref{table_modes} as the ``true'' ringdown signal.  These modes can be classified according to the following three criteria:
\begin{itemize}
    \item \textbf{Order in perturbation theory.} We consider both linear and quadratic modes. In fact, quadratic QNMs have recently been shown to be potentially detectable with LISA~\cite{Yi:2024elj}. Linear modes can be labeled with the angular and overtone numbers $\ell m n$.
    We choose to label quadratic modes by $(\ell_1 m_1 n_1)\times (\ell_2 m_2 n_2)$, where $\ell_i$ and $m_i$ are the angular numbers of the linear modes that source them. We will only consider quadratic modes with angular number given by $\ell = \ell_1+\ell_2$ and $m = m_1+m_2$. The frequencies and damping timescales of quadratic modes are given by $\omega_{\ell m n}=\omega_{\ell_1 m_1 n_1}+\omega_{\ell_2 m_2 n_2}$ and $\tau_{\ell m n}^{-1}=\tau_{\ell_1 m_1 n_1}^{-1}+\tau_{\ell_2 m_2 n_2}^{-1}$. 
    \item \textbf{Overtone number.} We only consider modes with $n = 0$ (fundamental modes) or  $n=1$ (overtone modes).  Higher overtone modes are typically too short-lived to be relevant for our analysis.
    \item \textbf{(Counter) rotation.} As shown in Eq.~\eqref{eq:QNMModelTimeDomain}, given a certain triplet $(\ell m n)$, there are two modes in the waveform with either positive or negative real frequency. We will refer as ``prograde'' to the modes with $\text{Sign}(\omega)=\text{Sign}(m)$, and as ``retrograde'' to those satisfying the opposite condition $\text{Sign}(\omega)=-\text{Sign}(m)$. Thus, each $\ell m n$ harmonic contains both a prograde and a retrograde mode.
    
\end{itemize}



Due to the homogeneity of the linear Teukolsky equation, the complex amplitudes of linear modes, unlike the frequencies, depend on the binary configuration before the merger. Therefore, they cannot be predicted by perturbation theory for a merger of comparable mass objects, although they can be extracted from NR simulations~\cite{London_2014,Cheung:2023vki, Pacilio_2024, Mitman:2025hgy}. We compute the true values of the amplitudes $A_{\ell m n}$ and phases $\Phi_{\ell m n}$ of the linear modes as functions of the masses $m_1,m_2$ and spins $a_1,a_2$ of the progenitors using the fits of~\cite{Cheung:2023vki} to NR simulations. 
To account for a starting time of the ringdown later than the luminosity peak, we multiply all complex amplitudes given in~\cite{Cheung:2023vki} by a suppression factor $e^{-i \omega_{\ell mn} t_0} e^{-t_0/\tau_{\ell mn}}$, which corresponds to evolving the model of Eq.~\eqref{eq:QNMModelTimeDomain} for a time $t_0$.

On the other hand, given the amplitude of linear modes, the amplitudes of quadratic modes can be computed in perturbation theory~\cite{Nakano:2007cj,Bucciotti:2023ets,Bucciotti:2024zyp,Bucciotti:2024jrv,Bucciotti:2025rxa,Ma:2024qcv,Khera:2024yrk}. We use the fits from NR simulations of~\cite{Khera:2024yrk} to obtain the spin-dependent amplitude of the quadratic modes that we include in our model. Notice that our assumption of equatorial symmetry implies that the ratio of quadratic to linear mode amplitudes is a single number independent of the initial conditions~\cite{Bucciotti:2024jrv,Khera:2024yrk}. 

The frequencies and damping times $\omega_{\ell m n}$ and $\tau_{\ell m n}$ are obtained by interpolating numerical values computed with the continued-fraction method~\cite{Berti:2005ys,Berti_2009,leaver}. 
However, as we will explain in the next section, for our analysis we need  derivatives of the template with respect to its parameters, which requires computing the derivatives of the QNMs frequencies with respect to the remnant spin. Only for this purpose, we resort to the analytic fits described in~\cite{Berti:2005ys}, instead of the more accurate interpolation. 
Those analytic fits are given by
\begin{equation}
\begin{split}
    &\omega_{\ell m n}=f_1+f_2(1-\chi_{\rm rem})^{f_3}\,;\\
    &Q_{\ell m n}=q_1+q_2(1-\chi_{\rm rem})^{q_3}\,;\\
    &\tau_{\ell m n}= \frac{2 Q_{\ell m n}}{\omega_{\ell m n}}\,,
\label{eq:freq_fit}
\end{split}
\end{equation}
where $\chi_{\rm rem}$ is the remnant spin, and $\{f_1,f_2,q_1,q_2,q_3\}$ are given in~\cite{Berti:2005ys}.
The mass and spin of the remnant, entering Eqs.~\eqref{eq:freq_fit} and ~\eqref{eq:ringdownModelFreqDomain}, are computed with the phenomenological formulae of~\cite{Barausse:2012qz, Hofmann:2016yih}.

\section{Assessing waveform inaccuracy}
\label{sec2}

We will now provide an overview of the general approach for assessing the impact of waveform inaccuracies in GW parameter estimation using the linear signal and Fisher approximations, based on~\cite{Cutler:2007mi,Cutler:1994ys,Flanagan:1997kp,Lindblom:2008cm}, and introduce the relevant notation and definitions.

Following~\cite{Cutler:2007mi}, we assume that there exists a template $h_{\rm GR}(\boldsymbol{\theta};t)$ with parameters $\boldsymbol{\theta}$ which, when evaluated at the true parameters $\boldsymbol{\theta} = \boldsymbol{\theta}_{\rm tr}$, can accurately describe the GW signal.
Geometrically, in the absence of noise, this simply means that the GW signal 
lies on the template manifold spanned by $h_{\rm GR}$~\cite{Cutler:2007mi,Lindblom:2008cm,Chandramouli:2024vhw}.
When using a less accurate template, denoted as $h_{\rm AP}$, even in the absence of noise, we would find that the best-fit parameters $\theta^i_{\rm bf}$ do not match the true parameters $\theta^i_{\rm tr}$, where $\theta^i$ are the parameters common to both $h_{\rm GR}$ and $h_{\rm AP}$.

The errors induced by such waveform inaccuracies are a type of \emph{systematic error}. 
Uncertainties in parameter estimation induced by detector noise are instead a type of \emph{statistical error}.
A waveform template is accurate when the systematic errors are smaller than the statistical errors.

In the following, for completeness, we review the derivation for the different errors, along with the waveform accuracy criteria.
We set the GW data in a given detector to be $d(t) = h_{\rm GR}(\boldsymbol{\theta}_{\rm tr};t) + n(t)$, where $n(t)$ is the detector noise.
For stationary and Gaussian noise, 
the likelihood $p(d|\boldsymbol{\theta})$ corresponding to the approximate template $h_{\rm AP}$ is then given, up to a constant, by
\begin{align}
    \log p\left(d \left| \right. \boldsymbol{\theta} \right)  = -\dfrac{1}{2} \left(d-h_{\rm AP} (\boldsymbol{\theta}) \left|\right. d-h_{\rm AP} (\boldsymbol{\theta})\right),\label{eq:log-likelihood}
\end{align}

For the parameters common to both $h_{\rm AP}$ and $h_{\rm GR}$, let $\theta_{\rm bf}^i$ be the best-fit parameters of $h_{\rm AP}$ that maximize the log-likelihood given by Eq.~\eqref{eq:log-likelihood}. 
We introduce the total error due to both waveform inaccuracy and detector noise as $\Delta \theta^i = \theta^i_{\rm bf} - \theta^i_{\rm tr}$. 
In the frequency domain, Gaussian and stationary noise is uncorrelated across frequencies~\cite{Cutler:1994ys}, i.e.,
\begin{equation} 
    \langle n(f)n^*(f')\rangle=\frac{1}{2}S_n(f)\delta_D(f-f')\,,
    \label{noise_2pf}
\end{equation}
where $\delta_D$ is the Dirac delta function and $S_n$ is the PSD.
The symbol $\langle\,\cdot\,\rangle$ indicates the ensemble average over all possible noise realizations.
For Gaussian and stationary noise, note that Eq.~\eqref{noise_2pf} also implies the completeness relation given by
\begin{equation}
    \langle\left( u |n\right)\left(n|v\right)\rangle = \left(u|v\right)\,,
    \label{completeness_relation}
\end{equation}
for two frequency-domain signals $u(f)$ and $v(f)$.
Given that stationary noise
is uncorrelated across frequencies,
we choose to work in the frequency domain to compute $\Delta \theta^i$.

In general, computing $\Delta \theta^i$ requires an exploration of the likelihood surface around the maximum likelihood point. 
For high-dimensional parameter spaces, this typically requires stochastic methods such as Markov Chain Monte Carlo or nested sampling.
However, when the SNR is large, and for small waveform modeling errors, one can simplify the computation of $\Delta \theta^i$ by using a combination of the Fisher approximation and the linear signal approximation.

We introduce the optimal SNR of the approximate template $h_{\rm AP}$ as
\begin{equation}
    \rho=\sqrt{\left(h_{\rm AP}\left|\right.h_{\rm AP}\right)}.
\end{equation} 
When $\rho \gg 1$, we can take advantage of the Fisher (or Laplace) approximation~\cite{Fisher:1922saa, Vallisneri:2007ev}. 
For the parameters that are common to $h_{\rm AP}$ and $h_{\rm GR}$, the likelihood region around $\theta_{\rm bf}$ is well described by a Gaussian distribution, and one can expand the template $h_{\rm AP}$ in terms of $\delta \theta^i = \theta^i - \theta^i_\mathrm{bf}$ up to leading order.
Explicitly, the likelihood then reads
\begin{equation}
    p(d|\boldsymbol{\theta}) \approx \mathcal{N} \exp \left[-\frac{1}{2}\Gamma_{ij}\delta\theta^i\delta\theta^j\right]\,,\label{eq:likelihood_fisher}
\end{equation}
where $\Gamma_{ij}$ is the Fisher Information Matrix (FIM), defined by 
\begin{equation}
    \Gamma_{ij}\equiv \left(\partial_i h_{\rm AP}\left.\right|\partial_j h_{\rm AP}\right)\left.\right|_{\theta=\theta_{\text{bf}}}\,,
\end{equation}
where we used the notation $\partial_i= \partial/\partial\theta^i$.
The normalization factor is given by $\mathcal{N} = \sqrt{\mathrm{det}\, \Gamma/(2\pi)}$.



The estimation of $\theta^i_{\rm bf}$ requires maximizing Eq.~\eqref{eq:log-likelihood}, which amounts to solving\footnote{Geometrically, the difference between the data and template is normal to the tangent subspace at the maximum-likelihood point~\cite{Cutler:2007mi,Chandramouli:2024vhw}.} 
\begin{equation}
    \left(\partial_j h_{\text{AP}}(\boldsymbol{\theta}_{\text{bf}})\left|\right.d-h_{\text{AP}}(\boldsymbol{\theta}_{\text{bf}})\right)=0\,, \label{eq:max-likelihood}
\end{equation}
where we have $d(f) = h_{\rm GR}(\boldsymbol{\theta}_{\rm tr};f)+n(f)$.
When the difference between $h_{\rm GR}$ and $h_{\rm AP}$ is small, we can also linearize $h_{\rm AP}$ in $\Delta \theta^i$, proceeding as in the linearization in $\delta \theta^i$.
By Taylor expanding $h_{\rm AP}$ around $\theta_{\rm bf}^i$ to linear order in $\Delta \theta^i$, inserting it into the waveform difference $d-h_{\rm AP}(\boldsymbol{\theta}_{\rm bf})$, and solving for $\Delta \theta^i$, we obtain
\begin{align}
    \begin{aligned}
        \Delta\theta^i &= \Delta^{(n)}\theta^i+\Delta^{\text{(Sys)}}\theta^i,\\
        \Delta^{(n)}\theta^i &=\left(\Gamma^{-1}\right)^{ij}\left(\partial_jh_{\text{AP}}\left|\right.n\right) \Big |_{\theta^i = \theta^i_{\rm tr}}, \\
        \Delta^{\text{(Sys)}}\theta^i &= \left(\Gamma^{-1}\right)^{ij}\left(\partial_jh_{\text{AP}}\left|\right.h_{\text{GR}}-h_{\text{AP}}\right) \Big |_{\theta^i = \theta^i_{\rm tr}}.
    \end{aligned}\label{eq:total_error}
\end{align}
In Eq.~\eqref{eq:total_error}, we have used the fact that in the linear signal approximation, 
the gradients of the waveform are identical at both $\theta_{\rm bf}^i$ and $\theta_{\rm tr}^i$, which also implies that $\Gamma^{ij}$ is the same when evaluated at $\theta^i_{\rm bf}$ or $\theta^i_{\rm tr}$.
The systematic error due to waveform inaccuracy is given by $\Delta^{\text{(Sys)}}\theta^i$.
Meanwhile, $\Delta^{(n)} \theta^i$ captures the bias due to a specific noise realization.
When averaged over all realizations, the one-point function $\langle \Delta^{(n)} \theta^i \rangle$ vanishes. 
In other words, averaging over all noise realizations is equivalent to setting the noise realization to zero~\cite{nissanke_2009,Vallisneri:2011ts}. 
However, the two-point function $\langle \Delta^{(n)}\theta^i \Delta^{(n)}\theta^j \rangle$ will not vanish, and, as we show below, quantifies the ensemble averaged statistical error.
We have that
\begin{equation}
\begin{split}
    \langle &\Delta^{(n)}\theta^i\Delta^{(n)}\theta^j\rangle= \\&=\langle\left(\Gamma^{-1}\right)^{ik}\left(\Gamma^{-1}\right)^{jl}\left(\partial_kh_{\text{AP}}\left|\right.n\right)\left(n\left|\right.\partial_lh_{\text{AP}}\right)\rangle\left.\right|_{\theta=\theta_{\text{tr}}} \\
    &=\left(\Gamma^{-1}\right)^{ij} \left.\right|_{\theta=\theta_{\text{tr}}}\,,
\end{split}
\end{equation}
where, in the second step, we exploited the completeness relation of Eq.~\eqref{completeness_relation} and the definition of FIM.
The statistical error is quantified through $\Delta^{(\rm St)} \theta^i \equiv \sqrt{\langle (\Delta^{(n)} \theta^i )^2\rangle}$, resulting in~\cite{Cutler:1994ys,Flanagan:1997kp,Vallisneri:2007ev}
\begin{equation}
    \Delta^{\text{(St)}}\theta^i = \sqrt{\left(\Gamma^{-1}\right)^{ii}}\,. 
    \label{eq:sys_bias_def}
\end{equation}

We define the waveform template $h_{\rm AP}$ to be accurate when
\begin{align}
    |\Delta^{\text{(Sys)}}\theta^i| < \Delta^{\text{(St)}}\theta^i, \label{eq:sys_stat_accuracy}
\end{align}
and likewise $h_{\rm AP}$ is inaccurate when the inequality in Eq.~\eqref{eq:sys_stat_accuracy} is violated.  
In our work, we use Eq.~\eqref{eq:sys_stat_accuracy} as a criterion to determine the minimum number of ringdown modes needed to ensure accurate parameter estimation.

Another way to quantify the accuracy of the waveform template $h_{\rm AP}$, relative to $h_{\rm GR}$, is through the \emph{match}, which is given by
\begin{equation}
\mathcal{M}=\max\limits_{t_c, \phi_c} \frac{(h_{\rm AP}|h_{\rm GR})}{\sqrt{(h_{\rm AP}|h_{\rm AP})(h_{\rm GR}|h_{\rm GR})}}.
\end{equation}
The match represents the normalized scalar product between the two templates, maximized over a relative time shift $t_c$ and phase shift $\phi_c$.
A value of $\mathcal{M}$ close to one indicates that the templates almost perfectly overlap. 
Note that one needs to align the templates by maximization over $\phi_c$ and $t_c$ so that any mismatch is entirely attributed to waveform inaccuracy and not to template misalignment.

Requiring that Eq.~\eqref{eq:sys_stat_accuracy} holds for all $\theta^i$ translates to~\cite{Chatziioannou:2017tdw,Chandramouli:2024vhw}
\begin{equation}
1-\mathcal{M}< \frac{D}{2\rho^2},\label{eq:mismatch_criterion}
\end{equation}
where $1-\mathcal{M}$ is the \emph{mismatch}, 
$D$ is the number of parameters of the model $h_{\rm AP}$, and $\rho$ is the optimal SNR. 
In the mismatch criterion given by Eq.~\eqref{eq:mismatch_criterion}, we have neglected the contribution from the fitting factor, which makes the criterion conservative~\cite{Flanagan:1997kp,Lindblom:2008cm,Chatziioannou:2017tdw,Purrer:2019jcp,Chandramouli:2024vhw,Toubiana:2024car}.
In other words, when Eq.~\eqref{eq:mismatch_criterion} is satisfied, the approximate waveform is sufficiently accurate.

Following~\cite{Moore:2018kvz}, we now discuss how to perform the maximization over $\phi_c$ and $t_c$ and compute $\mathcal{M}$.
Attributing the time and phase shift to the template $h_{\rm AP}$, i.e. defining $\hat{h}_{\rm AP}=e^{-i\phi_c+2\pi i f t_c}h_{\rm AP}$, the inner product to be maximized is
\begin{equation}
(h_{\rm GR}|\hat{h}_{\rm AP})=4\mathrm{Re}\left[e^{i\phi_c} \int_0^\infty \frac{h_{\rm GR}\,h^*_{\rm AP}}{S_n(f)}e^{-2 \pi i f t_c} df\right].
\end{equation}
We then define $\tilde{G}(f)=h_{\rm GR}\cdot h^*_{\rm AP}/S_n(f)$, so that
\begin{equation}
(h_{\rm GR}|\hat{h}_{AP})=4\mathrm{Re}[e^{i\phi_c}G(t_c)],\label{eqn:mismatch_numerator}
\end{equation}
where we used the definition of the inverse Fourier transform. 
The maximization of $(h_{\rm GR}|\hat{h}_{AP})$ over $\phi_c$ is carried out by simply taking the modulus of the complex quantity $G(t_c)$. 
The match then reduces to
\begin{equation}
\mathcal{M}=\mathcal{C} \cdot \max\limits_{t_c}|G(t_c)|,
\label{eq:mismatch}
\end{equation}
with the proportionality constant being $\mathcal{C}=4/\sqrt{(h_{\rm GR}|h_{\rm GR})(h_{\rm AP}|h_{\rm AP})}$.
To maximize the quantity in Eq.~\eqref{eq:mismatch} we employed the \texttt{minimize\_scalar} function from \texttt{scipy}, using the \texttt{`bounded'} method. This approach performs a minimization within specified bounds using Brent's algorithm, which does not require derivatives evaluation.



We end this section with a summary of the main points governing how we assess waveform accuracy: 
\begin{itemize}[left=0pt]
    \setlength \itemsep{0.25em}
    \item We use the Fisher and linear signal approximations to estimate the systematic error and statistical error when using the approximate template $h_{\rm AP}$. We then quantify the waveform to be accurate when Eq.~\eqref{eq:sys_stat_accuracy} is satisfied for every parameter in $h_{\rm AP}$.
    \item We also independently quantify the waveform accuracy using the mismatch criterion given by Eq.~\eqref{eq:mismatch_criterion}, which is valid for small mismatches and large SNR. Doing so allows us to further validate the use of Eq.~\eqref{eq:sys_stat_accuracy}, because the computation of the  mismatch does not explicitly use the linear signal approximation.
    \item Given the approximations made, our use of the waveform accuracy criteria (described by Eq.~\eqref{eq:sys_stat_accuracy} and Eq.~\eqref{eq:mismatch_criterion}) is conservative.
\end{itemize}

\section{Results}
\label{sec4}

\begin{figure}[h]
    \centering
    \includegraphics[width=0.7\textwidth, trim=4cm 0cm 1cm 0cm, clip]{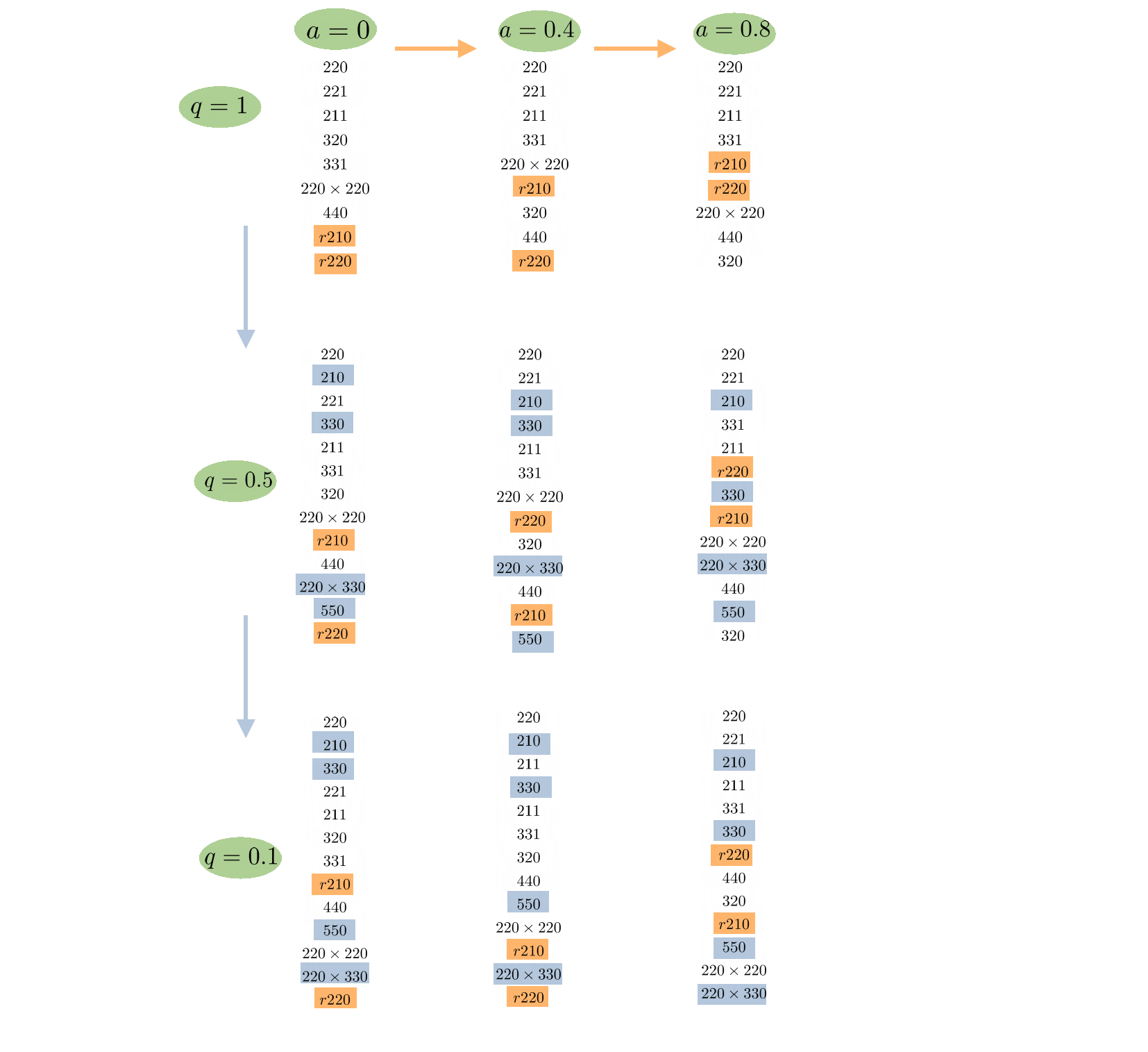}
    \caption{QNMs ordered by SNR from the loudest (up) to the quietest (down). We take 3 representative values of spin and mass ratio, fixing the primary mass to $10^6\,M_{\odot}$, the luminosity distance to $5 \text{Gpc}$, and the angles to $\theta = \psi=\iota=\pi/3$, $\phi=0$. The odd-$m$ modes are highlighted in blue. Note their absence in a mass-symmetric system, and their increasing relevance as we take a small $q$. Further observe that they tend to climb the ranking as we consider more spinning systems. 
    Besides these two features, for highly spinning and mass-asymmetric systems, overtones tend to become more important. The most peculiar case comes with the $331$, which becomes louder than the fundamental $330$.
    }
    \label{fig:Modes_Ordering}
\end{figure} 
\begin{figure}[t]
          \centering
\includegraphics[width =0.45\textwidth]{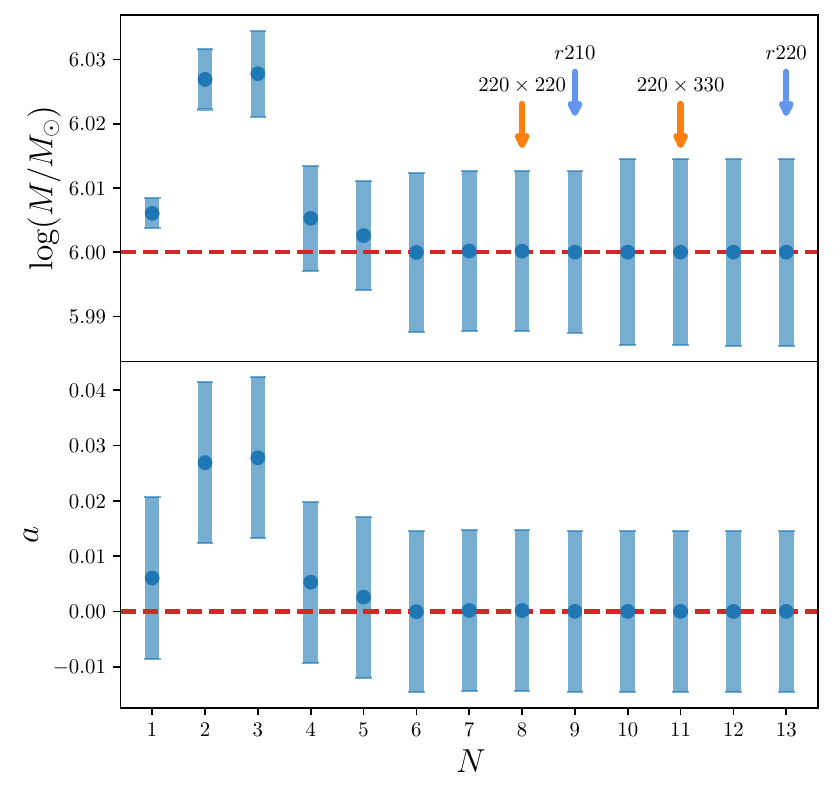}
 \caption{Comparison of statistical and systematic error for mass and spin, as a function of the number of modes $N$ for an approximate template $h_{\text{AP}}^{(N)}$. The selected system has  primary progenitor mass $10^6M_{\odot}$ and mass ratio $q = 0.5$ and it is located at a luminosity distance of $10 \,{\rm Gpc}$. The angles are fixed to the values $\theta = \psi = \iota = \pi/3$ and $\phi = 0$. The true value of the parameters is represented with the red dashed line, while the offset of the round points is given by the systematic error. Finally, the error bars represent the statistical error. 
The orange and blue arrows indicate respectively the points in which quadratic and linear retrograde modes are included.
Observe that the inclusion of more modes tends to tame the systematic error.
}
\label{fig:AvBias}
\end{figure}

\begin{figure*}[ht!]
    \centering
    \includegraphics[width=0.8\textwidth]{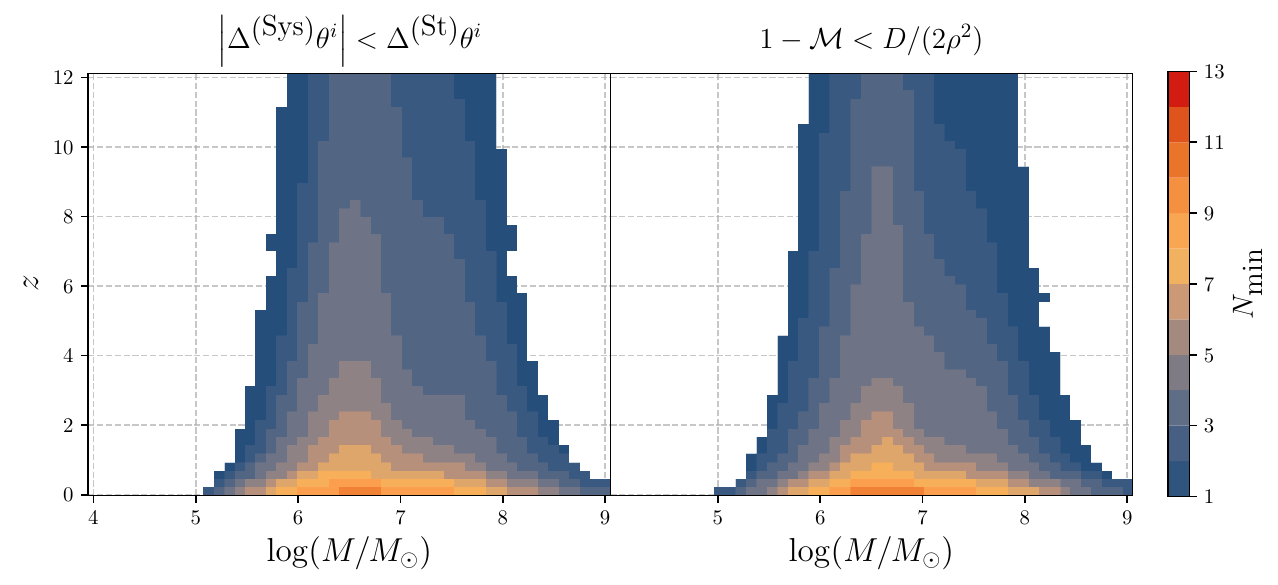}
    \caption{Minimum number of modes $N_{\text{min}}$, represented by the color code, as a function of primary mass and redshift. The mass ratio and the progenitor spins have been fixed to $q=0.5$ and $a_1=a_2=0$, while we have averaged over sky localization. 
    Observe that at low redshift $z<1$ and for $M/M_{\odot}\sim \mathcal{O}\left(10^{6-7}\right)$, we need $N_{\min} \in [8,10]$.
    Further note that the contours track the behavior of the SNR, given the PSD of LISA.
    }
    \label{fig:Waterfall}
\end{figure*}

Equipped with the formalism described in the previous sections, we now present the results of our work.

In the context of the ringdown, we will consider $h_{{\rm GR}}$ to be the template given by Eq.~\eqref{eq:ringdownModelFreqDomain} including all the 13 modes from the set described in Sec.~\ref{sec2c}. On the other hand, we will consider a family of approximate templates $h_{\rm AP}^{(N)}$ including a number $N<13$ of modes.
Recall that the set of parameters $\{\theta^i\}$ of $h_{\text{GR}}$ includes all the amplitudes and the phases of the $N$ modes, the logarithm of the remnant mass $\log M$, the spin of the remnant BH $\chi_{\rm rem}$, and the four angles $\theta$, $\phi$, $\psi$ and $\iota$, while we restrict ourselves to the case in which $\varphi = 0$, amounting to $D = 2(N+3)$ free parameters. Notice that we do not include the luminosity distance $d_L$ in the Fisher matrix, as it is completely degenerate with the mode amplitudes. Specifically, only the effective amplitudes $A_{\ell m n}^\text{eff}=A_{\ell m n}/d_L$ 
are measurable. Rather than working with $A_{\ell m n}^\text{eff}$, we choose to fix $d_L$ to its true value, in order to keep the amplitudes $\mathcal{O}(1)$. For the same reason, we do not include the angle $\varphi$, which has no impact on the observables besides shifting the mode phases $\phi_{\ell m n}$. In practice, we set $\varphi = 0$ for all  systems.


In order to study the dependence of the errors on the number of modes $N$, we adopt the SNR of single-mode template as a criterion for ordering the QNMs. 
The SNR of each mode, and the resulting hierarchy of the modes, are strongly dependent on the mass ratio and the spin. For instance, highly mass-symmetric systems result in faint odd-$\ell$ modes, while high spins can affect the spectrum in different ways, e.g. enhancing the contribution from retrograde modes. 
Additionally, the angles, in particular $\iota$, can influence the luminosity of the QNMs. 

In Fig.~\ref{fig:Modes_Ordering}, we schematically show the dependence of the QNMs on the mass ratio and spin. 
Crucially, we note that 
the overtone modes, described by the fits of~\cite{Cheung:2023vki},
tend to become louder and dominate the spectrum when we consider systems with large spin and small mass ratio. 
We attribute this behavior to the extrapolation of the amplitude values to very early time, where the fitted amplitude is still unstable. 
Hence, in order to get physically robust results, one should carefully choose the ringdown starting time, which could in principle vary for different systems. 
In our work, we adopt $t_0=20 M$ as a conservative choice for systems with non-spinning progenitors, motivated by a comparison with NR waveforms that we describe in more detail in Appendix \ref{NR_comparison}.

Given the ordering procedure described above, we proceed to compute the systematic errors $\Delta^{(\rm Sys)}\theta^i$ and statistical errors $\Delta^{(\rm St)}\theta^i$, as described in Sec.~\ref{sec2}\footnote{We also note here that while the formalism in Sec.~\ref{sec2} is presented for a single detector, we apply it for the LISA detector network described in Sec.~\ref{sec3}. We do so by summing over the relevant inner products  to obtain the network statistical and systematic errors (see Appendix A of~\cite{Kapil:2024zdn}).}.
In order to compute $\Delta^{(\rm Sys)}\theta^i$ and $\Delta^{(\rm St)}\theta^i$, we incorporate
Gaussian priors for the parameters of $h_{\rm AP}$, following~\cite{Cutler:1994ys,Vallisneri:2007ev,Favata:2013rwa}.
Specifically, we implement these priors by 
shifting the FIM by a diagonal matrix $\varepsilon_{ij}$, namely $\Gamma_{ij}\rightarrow \Gamma_{ij}+\varepsilon_{ij}$.
Concretely, one  can show that $\varepsilon_{ii} = 1/\sigma^2_i$, where $\sigma_i$ is the standard deviation of the Gaussian prior for parameter $\theta^i$.
A more informative prior on $\theta^i$ corresponds to a larger $\varepsilon_{ii}$.
Due to the large dimensionality of our FIMs, the inclusion of these terms can also be interpreted as a way of conditioning them for inversion.
In practice, we use a prior $\varepsilon_{ii}=0.01$ for amplitudes and phases of the modes, and $\varepsilon_{ii}=0.1$ for the logarithms of  mass and  spin. This choice corresponds to a Gaussian prior with a width $\sigma_i = 10$ for amplitudes and phases and $\sigma_i = 3.16$ for logarithm of the mass and spin. Amplitudes and phases are typically $\sim \mathcal{O}(1)$ or less, the spin ranges in the interval $[0,1)$, and the error on the mass scales with the mass itself, hence, these priors are uninformative and do not affect the results. 
For the angles, we set $\varepsilon_{\theta\theta}=\varepsilon_{\iota\iota}=0.1$ and $\varepsilon_{\phi\phi}=\varepsilon_{\psi\psi}=0.025$. This corresponds to covering the whole celestial sphere and all the possible values of inclination and polarization angle. 

In Fig.~\ref{fig:AvBias}, we show the systematic and statistical errors for mass and spin, computed with Eqs.~\eqref{eq:total_error} and~\eqref{eq:sys_bias_def}.
We fixed the primary progenitor mass to $10^6M_{\odot}$, the mass ratio $q$ to 0.5 and the spins to $a_1=a_2=0$. The angles are fixed to $\theta=\psi=\iota =\pi/3$ and $\phi = 0$ and the luminosity distance to $10\,{\rm Gpc}$.
The systematic error can be understood as the difference between the best-fit values, depicted by the blue circles, and the true values, represented by red dashed lines.  The statistical error is presented as the blue vertical error bars for each setup. We show both quantities as functions of the number of modes included in the approximate template.

When a large number of modes are included, one can clearly observe that the systematic bias is mitigated.
To understand the trend of the systematic bias with $N$, we first discuss the behavior of the statistical error with increasing $N$.
Here, there are, in fact, two competing effects at play -- one due to the increasing dimensionality and another due to the increasing SNR.
Adding new modes increases the dimensionality of the parameter space, which contributes to an increase in the statistical error.
At the same time, the increase in SNR with increasing $N$ has the opposite effect.
Overall, the increase in dimensionality is the dominant contribution when $N$ is large, owing to a decreasing SNR contribution from the higher modes. 

We now turn to the behavior of the systematic error as a function of $N$.
The early trend  for $N<3$ is oscillatory.
In Appendix~\ref{Linear_regime}, we show that this behavior for small $N$ corresponds to the non-perturbative regime, where the linear signal expansion is not valid, owing to the large dephasing between $h_{\rm GR}$ and $h_{\rm AP}^{(N)}$.
As we include more modes, specifically when $N>3$, we see a clear trend of decreasing systematic error  with increasing $N$.
Moreover, when $N >3$, we observe that the systematic bias is negligible, as the true values are contained within the statistical errors.

Given the behavior of the systematic and statistical errors with $N$ for a specific system, we now move on to discuss what is the minimum number of modes $N_{\min}$ needed for unbiased parameter estimation across parameter space. 
We independently compute $N_{\min}$ using the two criteria of Eq.~\eqref{eq:sys_stat_accuracy} and Eq.~\eqref{eq:mismatch_criterion}.
Given the non-trivial behavior of both the systematic and statistical errors with $N$, we implement the following scheme to ensure that our estimate of $N_{\min}$ is robust.
We start with $h_{\rm AP}^{(1)}$, which includes only the loudest $220$ mode and evaluate the criterion given by Eq.~\eqref{eq:sys_stat_accuracy}. 
If the criterion is already satisfied, we estimate  $N_{\min}=1$.
However, to ensure that this estimate is robust, we add the next two modes, and check whether the criterion is still satisfied.
If instead the criterion is not satisfied for $h_{\rm AP}^{(1)}$, we move on to $N=2$ modes, and repeat the above steps until we find the $h_{\rm AP}^{(N)}$ that satisfies the criterion.


We first focus on the dependence of $N_{\min}$ on mass $M$ and redshift $z$, by fixing the spins to $a_1=a_2=0$, and the mass ratio to $q = 0.5$. We show the dependence on $a_1,a_2$, and $q$ in Appendix~\ref{app_spins}.
To compute $N_{\min}$ as a function of $(M,z)$, we average over several configurations of sky localization, polarization and inclination angles. 
In practice, we generate a large number of random configurations \footnote{The number of configurations is chosen so that the error on the Monte Carlo average is under control, typically $\delta N_{\rm min}\ll1$.} of $(\theta, \phi, \psi, \iota)$ and compute the Monte Carlo-averaged $N_{\text{min}}$ in both cases.
Afterwards, we smooth out the sampling fluctuations 
n the data with a moving average algorithm in two dimensions. 
The final result is then rounded to the closest integer.

In the left panel of Fig.~\ref{fig:Waterfall}, we show the resulting $N_{\min}$ obtained using Eq.~\eqref{eq:sys_stat_accuracy} (within the linear-signal approximation for the systematic and statistical errors). The dependence of $N_\text{min}$ on the mass and redshift tracks the SNR behavior, controlled by the LISA PSD. At small redshifts $z<1$ and for masses $M/M_\odot \sim \mathcal{O}(10^{6-7})$, i.e. in the  most sensitive part of the LISA frequency band, a number of modes in the range $N_\text{min} \in [8,10]$ will be needed to ensure an unbiased estimate of the parameters. At higher redshift, we find that there is a wide region of  parameter space where $N_{\text{min}} \in [3,6]$ is required. For a given mass, increasing the redshift results in a fainter signal, for which fewer modes are required. Similarly, for either very high or low mass, even at low redshift, $N_\text{min}$ decreases. 
We checked that if we impose Eq.~\eqref{eq:sys_stat_accuracy} only for
the intrinsic parameters (mass, spin, amplitudes and phases), rather than for all parameters (intrinsic and extrinsic), the value of $N_{\text{min}}$ is essentially unchanged.
 
In the right panel of Fig.~\ref{fig:Waterfall}, we show $N_{\min}$ as obtained using the mismatch criterion given by Eq.~\eqref{eq:mismatch_criterion}. 
We find strong agreement between the results obtained using the two criteria.
The main reason for this, as discussed in Sec.~\ref{sec3}, is that when Eq.~\eqref{eq:sys_stat_accuracy} is satisfied for every parameter, then Eq.~\eqref{eq:mismatch_criterion} is also satisfied.
The main difference between the two criteria is that the linear signal approximation is explicitly used in Eq.~\eqref{eq:sys_stat_accuracy}, while it is used implicitly in Eq.~\eqref{eq:mismatch_criterion}.
Specifically, we did not use the linear signal approximation in computing $\mathcal{M}$, which results in slightly different estimates for $N_{\min}$ in different parts of the parameter space. 
Overall, we observe that in most of the parameter space, the mismatch criterion results in a slightly lower estimate of $N_{\min}$.

We finally stress that the FIM approach relies on the assumption of high SNR. This requirement is fulfilled by typical massive black hole binaries in the LISA band, which span a wide range of the parameter space shown in Fig.~\ref{fig:Waterfall}. 
Sources with SNR $\in (8,15)$ correspond to the boundaries of Fig.~\ref{fig:Waterfall}.
Our FIM approach thus is not reliable for this (small) region of parameter space with low SNR~\cite{Vallisneri:2007ev,Nicholson:1997qh,Rodriguez:2013mla,Cokelaer:2008zz}.
For these sources, a full Bayesian analysis would be needed to validate the estimation of $N_{\min}$ against our FIM-based prediction.

\section{Conclusions}
In this work, we carried out an analysis of the systematic bias due to an incomplete description of the GW ringdown signal, as observed by LISA. We employed a frequency-domain waveform template with a maximal set of 13 modes including linear QNMs--considering both fundamental modes and overtones--and two quadratic QNMs. 
We ranked the QNMs by their single-mode  SNR, and superimposed them to construct
approximate ringdown templates of increasing accuracy.

Our main result, represented in the waterfall plot of Fig.~\ref{fig:Waterfall}, is the minimum number of modes $N_{\min}$ required for unbiased parameter estimation.
As discussed in Sec.~\ref{sec2}, due to the high-frequency cutoff introduced in our frequency-domain model to mitigate spectral leakage from the dominant mode, the estimate for \( N_\text{min} \) should be interpreted as a lower bound that is insensitive to the choice of time window, used to isolate the ringdown stage.
We determined $N_{\min}$ based on two equivalent waveform accuracy criteria: a direct comparison of systematic and statistical errors within the linear-signal approximation, and, separately, via the mismatch criterion given by Eq.~\eqref{eq:mismatch_criterion} (with both methods in excellent agreement). 
Our findings highlight the importance of carefully selecting the waveform template across the LISA parameter space. 
For sources at redshifts around $z \sim 2$--$6$, corresponding to the peak in the merger rate of massive BH binary formation and coalescence models~\cite{LISA:2024hlh}, we find that a minimum of $3$--$6$ modes are needed to avoid systematic biases. This result holds if the modes are included in the template following a single-mode SNR ranking, starting from the loudest for a given system. Note that the presence of each mode depends on the excitation amplitude in the pre-ringdown phase, therefore, the observation of the QNMs is highly linked to the configuration of the system.
For massive BH binary sources at low redshifts of $z<1$, as many as $10$ modes may be required to ensure unbiased parameter estimation. 

While our work provides a robust and conservative estimate of the minimum number of QNMs required for controlling systematics, there are some limitations that should be acknowledged. 
First, we assume that the true GR signal is a pure linear superposition of QNMs, as given by the waveform in Eq.~\eqref{eq:QNMModelTimeDomain}. This representation is valid only within an intermediate time window, as it excludes the prompt response at early times and the power-law tails at late times. Although our high-frequency cutoff mitigates spectral leakage due to the windowing, a
full Bayesian \textit{time-domain} analysis is an important future
direction to further assess the robustness of our results.

Second, while there is general agreement across different mode extraction algorithms for the fundamental modes, we observed some discrepancies in the overtone amplitudes predicted by different fits~\cite{Cheung:2023vki, MaganaZertuche:2021syq, MaganaZertuche:2024ajz, Pacilio_2024} for certain combinations of source parameters. This could be a source of systematic errors {\it per se}, especially for systems with loud overtones.

We would also like to emphasize that, while the use of the low-frequency approximation --- compared to the full LISA response --- may have a limited effect on the general results, the omission of Time Delay Interferometry (TDI) can have a larger impact. This is particularly relevant for low-mass sources, whose ringdown phase may be shorter than the light travel time across the LISA arms. In general, the TDI response overlaps signals from different times, mixing pre-merger, merger and ringdown components. In order to isolate the ringdown phase and mitigate this issue, the analysis could be started at times later than approximately 68 s~\footnote{This corresponds to the light travel time for a second-generation TDI configuration.} after the luminosity peak. However, this delay can span the entire ringdown duration for some low-mass sources. Therefore, new methods for analyzing ringdown signals are needed to account for complex detector responses, such as that of LISA.


\section*{Acknowledgments}
We warmly thank Emanuele Berti, Mark Cheung, Francesco Crescimbeni and Sophia Yi for insightful discussion on preliminary results. We also thank Zexin Hu for pointing out the need for a high-frequency mass-dependent cutoff.
The authors acknowledge support from the European Union’s H2020 ERC Consolidator Grant ``GRavity from Astrophysical to Microscopic Scales'' (Grant No. GRAMS-815673),  the European Union’s Horizon  
ERC Synergy Grant ``Making Sense of the Unexpected in the Gravitational-Wave Sky'' (Grant No. GWSky-101167314), the PRIN 2022 grant ``GUVIRP - Gravity tests in the UltraViolet and InfraRed with Pulsar timing'', and the EU Horizon 2020 Research and Innovation Programme under the Marie Sklodowska-Curie Grant Agreement No. 101007855. A.K. acknowledges funding from the FCT project ``Gravitational waves as a new probe of fundamental physics and astrophysics'' grant agreement 2023.07357.CEECIND/CP2830/CT0003. C.P. and E.B. acknowledge support from Agenzia Spaziale Italiana (ASI), Project No.~2024-36-HH.0, ``Attività per la fase B2/C della missione LISA''.  
\appendix

\section{Reliability of the linear signal regime}
\label{Linear_regime}

In Sec.~\ref{sec2}, we used the linear signal approximation to estimate both, statistical and systematic errors.
For an accurate estimation of the statistical error, the validity of the linear signal approximation coincides with the validity of the Fisher approximation~\cite{Vallisneri:2007ev}.
Typically, when the SNR is large, considering linear deviations in the template around the maximum likelihood is equivalent to approximating the likelihood surface around the maximum as Gaussian.

We also used the linear signal approximation to estimate $\Delta^{(\rm Sys)} \theta^i$, which is the systematic error caused by the waveform inaccuracy. 
Essentially, we made the approximation that the difference between $h_{\rm AP}(\boldsymbol{\theta}_{\rm bf};f)$ and $h_{\rm AP}(\boldsymbol{\theta}_{\rm tr};f)$ is linear in $\Delta^{(\rm Sys)} \theta^i$.
The validity criteria of the linear signal approximation for estimating statistical errors (as given in~\cite{Vallisneri:2007ev}) does not apply for estimating systematic errors (which are induced by waveform inaccuracy).
A fundamental reason for this is that the systematic error is independent of the SNR, which can be seen from the fact that $\Delta^{(\rm Sys)} \theta^i$ given Eq.~\eqref{eq:total_error} is SNR scale invariant.
The SNR scale invariance of the systematic error holds even beyond the linear signal approximation because Eq.~\eqref{eq:max-likelihood} is SNR scale invariant, making the estimation of $\theta^i_{\rm bf}$ (which is a solution to Eq.~\eqref{eq:max-likelihood}) also SNR scale invariant.
The validity of the linear signal approximation for computing the systematic error is essentially tied to the magnitude of difference between the two waveform templates.
When the difference between $h_{\rm GR}$ and $h_{\rm AP}$ is ``large'', we expect that the simple linear approximation should not hold.

To further quantify the validity regime of the linear signal approximation, we obtain the explicit dependence of $\Delta \theta^i$ on the template difference by considering linear changes between $h_{\rm GR}$ and $h_{\rm AP}$.
Following~\cite{Cutler:2007mi,Chandramouli:2024vhw}, we first express $h_{\rm GR}(\boldsymbol{\theta}_{\rm tr};f) = A_{\rm GR}(\boldsymbol{\theta}_{\rm tr};f) \exp\left[ i\Psi_{\rm GR}(\boldsymbol{\theta}_{\rm tr};f)\right]$ and $h_{\rm AP}(\boldsymbol{\theta}_{\rm tr};f) = A_{\rm AP} (\boldsymbol{\theta}_{\rm tr};f)\exp\left[ i\Psi_{\rm AP}(\boldsymbol{\theta}_{\rm tr};f)\right]$, where recall that $h_{\rm AP}$ is only evaluated at the parameters common with $h_{\rm GR}$.
Introducing $\Delta A(\boldsymbol{\theta}_{\rm tr};f) \approx A_{\rm GR}(\boldsymbol{\theta}_{\rm tr};f) - A_{\rm AP}(\boldsymbol{\theta}_{\rm tr};f) $, and likewise $\Delta \Psi(\boldsymbol{\theta}_{\rm tr};f) \approx \Psi_{\rm GR}(\boldsymbol{\theta}_{\rm tr};f) - \Psi_{\rm AP}(\boldsymbol{\theta}_{\rm tr};f) $, the systematic error $\Delta^{(\rm Sys)}\theta^i$ simplifies to
\begin{align}
    \Delta^{(\rm Sys)}\theta^a \approx \underbrace{\left( \Gamma^{-1} \right)^{ab} \left( \left[\Delta A + i A_{\rm AP} \Delta \Psi  \right] e^{i \Psi_{\rm AP}} \Big | \partial_b h_{\rm AP} \right)}_{\theta^a = \theta^a_{\rm tr}}.
\end{align}
Since $\Delta^{(\rm Sys)}\theta^a$ scales linearly with both $\Delta A$ and $\Delta \Psi$, neglecting $\mathcal{O}((\Delta^{(\rm Sys)}\theta^a)^2)$ is formally equivalent to neglecting $\mathcal{O}((\Delta \Psi)^2)$, $\mathcal{O}((\Delta A)^2)$, and $\mathcal{O}((\Delta \Psi)(\Delta A))$ terms.
Thus, the linear regime is determined by $\Delta \Psi \ll 1$ and $\Delta A \ll A_{\rm AP}$ (or $\Delta A \ll A_{\rm GR}$).
For instance, when $\Delta \Psi \gtrsim 1$, we expect significant dephasing between the two templates, resulting in large biases and a breakdown of the linear signal approximation. 
The amplitude corrections are small when the relative waveform error $(\delta h | \delta h)/\rho^2$ is small, where $\delta h = h_{\rm GR} - h_{\rm AP}$. 
Indeed, when the relative waveform error is small and the SNR ratio of $h_{\rm GR}$ and $h_{\rm AP}$ 
is close to 1, we also have that~\cite{Flanagan:1997kp,Lindblom:2008cm,Chandramouli:2024vhw,Read:2023hkv} $(\delta h | \delta h) /(2\rho^2) \approx (1-\mathcal{M})$.
In the mismatch criterion given by Eq.~\eqref{eq:mismatch_criterion}, we essentially neglect relative waveform and relative SNR errors, and work in the limit of the match being close to 1.
Thus, by checking when $(\delta h | \delta h)/\rho^2$, we validate the linear signal approximation, which is explicitly used in Eq.~\eqref{eq:total_error}, and implicitly used in Eq.~\eqref{eq:mismatch_criterion}. 
\begin{figure}[t]
    \centering
    \includegraphics[width =0.4\textwidth]{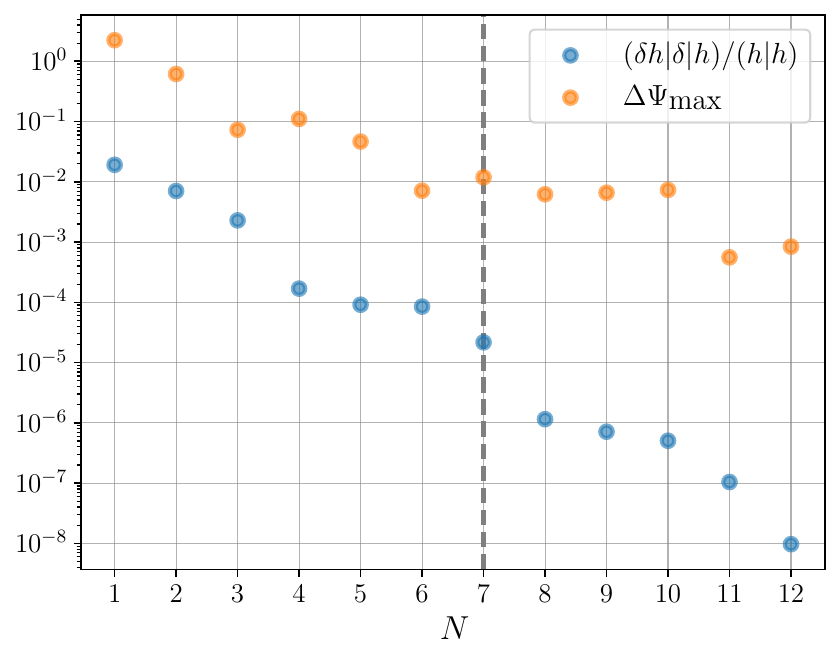}
    \caption{Relative waveform error (blue dots) and the maximum dephasing (orange dots) between $h_{\text{AP}}$ and the fiducial template, as a function of the number of modes. The system has primary mass $5\times 10^6\,M_{\odot}$ and is located at luminosity distance $5 \,{\rm Gpc}$, with fixed $\theta=\psi=\iota = \pi/3$ and $\phi = 0$. $N_{\text{min}}= 7$ is shown with the vertical dashed line.
    Observe that with increasing $N$, the dephasing and relative waveform errors decrease appreciably, allowing for the use of the linear signal approximation in estimating $N_{\min}$.
   }
    \label{fig:Check_Linearity}
\end{figure}

In Fig.~\ref{fig:Check_Linearity}, using the fiducial ringdown template described by Eq.~\eqref{eq:ringdownModelFreqDomain}, we illustrate the dependence of the maximum dephasing $\Delta \Psi_{\max}$ and the relative waveform error $(\delta h | \delta h)/\rho^2$ on the number of modes $N$.
We consider the same system as the one in Fig.~\ref{fig:AvBias}, where we fixed $\theta=\psi=\iota = \pi/3$ and $\phi = 0$.
Recall that we estimated the minimum number of modes for this system to be $N_{\min} = 7$.
As expected, with increasing $N$, both $\Delta \Psi_{\max}$ and $(\delta h | \delta h)/\rho^2$ decrease, implying that the absolute and relative errors are increasingly negligible when more modes are included.

We see that when $N < 3$, although relative waveform errors are at the percent level, the maximum dephasing is close to 1, suggesting that the linear signal approximation is not valid.
Recall that one of the features that we observed in Fig.~\ref{fig:AvBias} was the oscillatory behavior of $\Delta^{(\rm Sys)}\theta^i$ with increasing $N$ for $N<3$.
Such an oscillatory behavior is tied to the breakdown of the linear signal approximation. 
We can see this through the oscillatory $\exp(i\Delta \Psi)$ dependence in $\Delta^{(\rm Sys)} \theta^i$ (when one does not further linearize $\Delta^{(\rm Sys)} \theta^i$ in $\Delta \Psi$).
As $N$ increases to $3\leq N < 6$, we observe that $\Delta \Psi_{\max}$ drops to $\mathcal{O}(10^{-1})$, and we enter the regime where the linear signal approximation is valid. 
Crucially, for $N\geq 6$, $\Delta \Psi_{\max}$ is $\mathcal{O}(10^{-2})$, ensuring that the linear signal approximation is indeed valid in estimating the minimum number of modes, and that our estimate of $N_{\min} = 7$ is robust.
While we demonstrated this validity for a particular system in Fig.~\ref{fig:Check_Linearity}, the arguments apply across parameter space. 
The agreement between our estimate of $N_{\min}$ based on the (explicit linear signal approximation) criteria in Eq.~\eqref{eq:sys_stat_accuracy} and the (implicit linear signal approximation) criteria in Eq.~\eqref{eq:mismatch_criterion} further strengthens the robustness and self-consistency obtained within the linear signal approximation.

\section{Comparison with NR waveforms}
\label{NR_comparison}


When fitting numerical simulations using a superposition of QNMs, the analysis typically begins at progressively later times until the linear regime associated with a particular QNM emerges within the nonlinear numerical solution~\cite{Cheung:2023vki, MaganaZertuche:2024ajz, Pacilio_2024}; see, however,~\cite{Mitman:2025hgy} for a different perspective. Thus, one natural approach to assess the presence of a QNM consists in observing a stable amplitude for a given time duration.
This generally occurs at intermediate times, with the exception of the dominant mode $220$. Once the amplitude is stable, its value is extracted and extrapolated to t$_{peak}$, with t$_{peak}$ the time of the luminosity peak of the $(2,2)$ mode.
Therefore, highly damped QNMs, such as the overtones, present fitted amplitudes much larger than actually present in the numerical solution. This means that if one were to compute the SNR of the individual QNMs close to $t=t_\text{peak}$, the SNR of the overtones would be larger than for the fundamental tones.

As explained in Sec.~\ref{sec2}, to prevent this effect and, more generally, back-extrapolation from corrupting our analysis, we choose the ringdown starting time $t=t_0$ by comparing the ringdown model in Eq.~\eqref{eq:ringdownModelFreqDomain} (with the fits in~\cite{Cheung:2023vki,Khera:2024yrk}) directly with NR simulations. In particular, we require that, for each $(\ell,m)$, the difference between the amplitude obtained from the fits and that extracted from the simulation to be smaller than a prescribed tolerance. We select two sources from the catalog of the SXS collaboration~\cite{Boyle:2019kee}, SXS:BBH:0303 and SXS:BHB:0257, which correspond to ($a_1=a_2=0, \, q=0.1$) and ($a_1=a_2=0.85, \, q=0.5$) respectively. To compare the amplitudes, we first decompose the NR signal into its multipole $(\ell,m)$ components and then compare them with $h_{\ell m}^{Fit}$, obtained from the fitted amplitudes after summing over the overtone number, i.e. 


\begin{figure}[b]
    \centering
    \begin{subfigure}[b]{0.48\textwidth}
        \includegraphics[width=\linewidth]{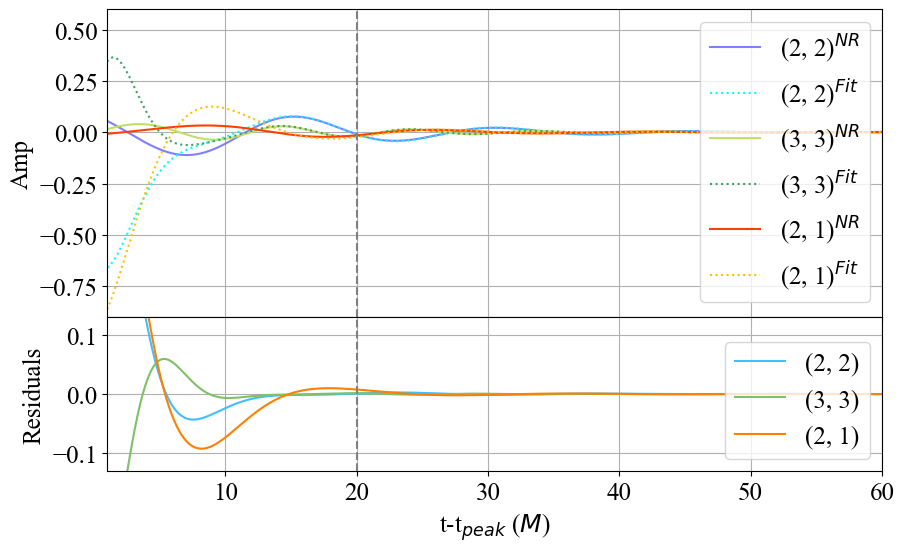}
        \caption{SXS:BBH:0303 with non-rotating BHs and q=0.1.}
        \label{fig:NR_0303}
    \end{subfigure}
    \begin{subfigure}[b]{0.48\textwidth}
        \includegraphics[width=\linewidth]{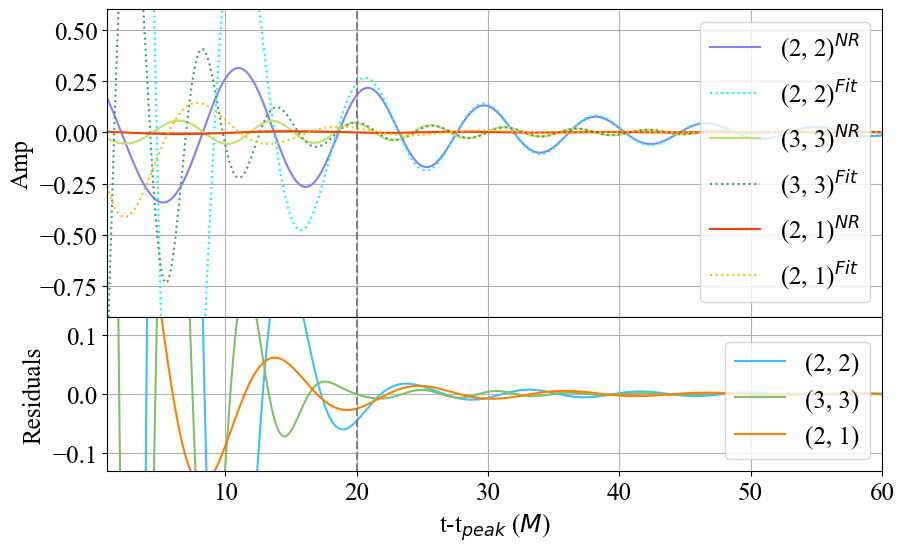}
        \caption{SXS:BBH:0257 with $a_1 = a_2=0.85$ and $q=0.5$.}
        \label{fig:NR_0257}
    \end{subfigure}
    \caption{\label{fig:NR_comparison} Comparison of multipoles drawn from NR simulations with multipoles obtained from Eq.~\eqref{eq:fit_multipoles}.}
     
\end{figure}

\begin{equation}
h_{\ell m}^{Fit} = \sum_{\ell’=-\ell}^{\ell’=\ell} \mu_{\ell\ell’m’n’} h_{\ell’m’n’} \delta_{m m’},
\label{eq:fit_multipoles}
\end{equation}

\begin{figure*}[t]
    \centering
    \includegraphics[width=0.8\textwidth]{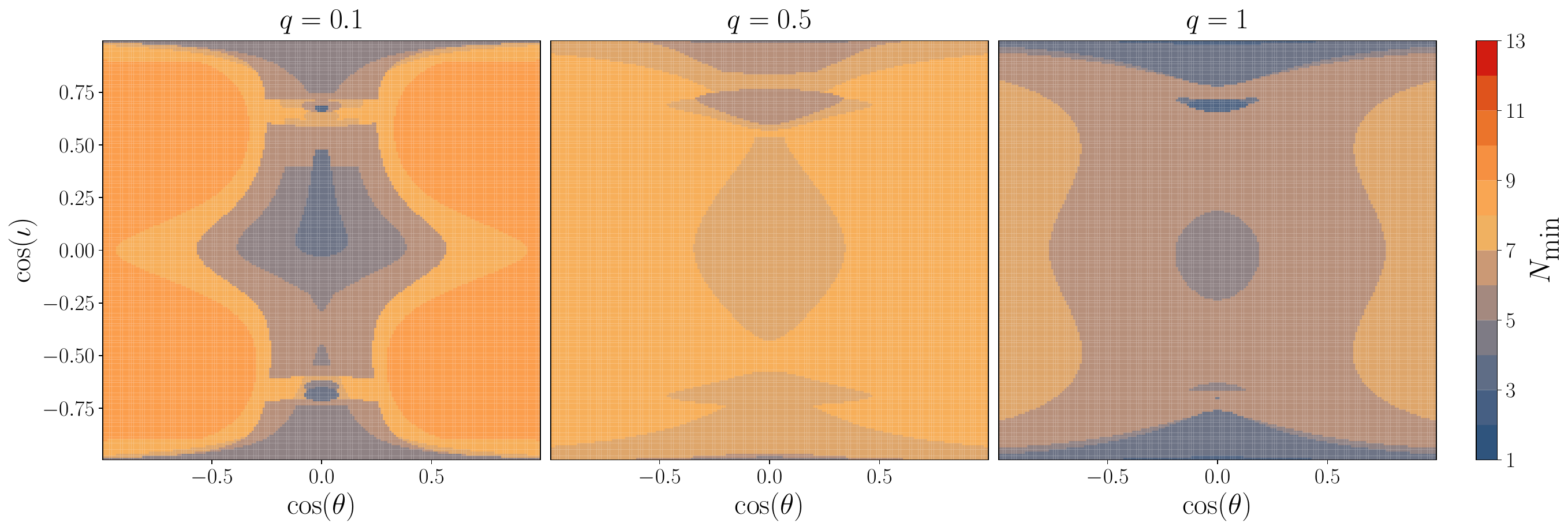}
    \caption{Dependence of $N_{\text{min}}$ on the cosine of the angles $\theta$ and $\iota$, for three values of the mass ratio of the progenitors $q$ and $a_1=a_2=0$. The primary BH mass and the luminosity distance have been fixed respectively to $5 \times 10^6 M_{\odot}$ and $5\,\text{Mpc}$. $N_{\text{min}}$ is represented by the color code. Observe that the color pattern is symmetric with respect to $\cos\theta=0$, while the corresponding symmetry about $\cos\iota=0$ is slightly broken. Observe also that some patterns  present in the right panel disappear in the middle one, to appear again in the mass-symmetric case (left panels). This is the consequence of the competing effects at play when the progenitor mass ratio changes, namely (de)excitation of odd-$m$ modes against changes in the global SNR.}
    \label{fig:anlge_dep}
\end{figure*}

where $\mu_{\ell \ell’m'n'}$ are the spherical-spheroidal mixing coefficients~\cite{Berti_Klein, London_2019}. 
Note that in this expression, the linear perturbation framework is respected and only the modes that fulfill $m’=m$ are considered. To give some context, modes with $m’\neq m$ (known as ‘recoil’ modes) are present in several NR waveforms due to extrapolation at null infinity. If, instead, one extracts the numerical waveforms with the Cauchy-characteristic extraction technique in the superrest (BMS) frame~\cite{CCE, Mitman:2022kwt, MaganaZertuche:2021syq}, these recoil modes vanish.

The amplitude residuals and the two waveforms are shown in Fig.~\ref{fig:NR_comparison}.
Note that at early times, the overtones from the fitted model dominate the signal, while matching the NR waveform as the time increases. Considering two different sets of parameters, we decide to set the starting time of the ringdown to $t_{0} = 20M$ as a compromise between the residuals and the loss of SNR in both cases.

\section{The impact of spins, mass ratio and angles on $N_{\text{min}}$}

\label{app_spins}
In this Appendix, we comment more extensively on the dependence of $N_{\text{min}}$ on the other parameters, which we briefly discussed in the main text. We restrict the discussion to binary systems with fixed primary BH mass of $5\times10^6 M_{\odot}$ at luminosity distance $d_L=5 \text{Mpc}$.

In Fig.~\ref{fig:anlge_dep}, we show the dependence of $N_{\text{min}}$ on the angles $\iota$ and $\theta$, fixing the progenitors spins to $a_1=a_2=0$ and $q =0.1,\,0.5,\,1$. First, we notice that the patterns are completely symmetric around $\cos\theta =0$, whereas there is a slight symmetry breaking in the two hemispheres defined by $\cos\iota=0$. This is indeed what we expect, as two configurations with opposite sky localization with respect to the horizon are completely equivalent, while the corresponding symmetry in the BH frame is broken by the direction of the spin. Furthermore, it can be observed that the case with $q=0.5$ represents an optimal configuration, where more modes are needed on average, compared to the other two cases. This happens because, as we decrease the mass ratio starting from $q=1$, the QNM spectrum becomes richer, as odd-$m$ modes are switched on. On the other hand, as we go to smaller mass ratios while keeping the primary mass fixed, the size of the region in which $N_{\text{min}}<5$ increases again. This happens because we are decreasing the total SNR of the event, although many modes are excited. For $q=0.1$ there are still, however, marginal regions in which $N_{\text{min}}>6$.

\begin{figure}[!t]
    \centering
    \includegraphics[width =0.4\textwidth]{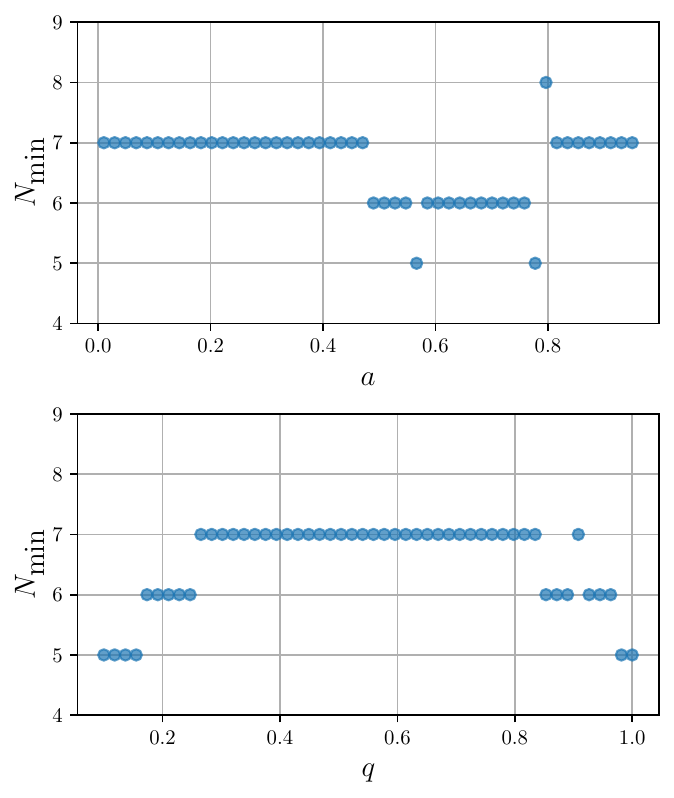}
    \caption{Dependence of $N_{\text{min}}$ on progenitor spins and mass ratio. The primary BH mass and the luminosity distance have been fixed respectively to $5 \times 10^6 M_{\odot}$ and $10\,\text{Gpc}$, while the angles have been fixed to $\theta=\psi=\iota=\pi/3$, $\phi=0$. The jumps correspond to crossing of two or more modes in the single-mode SNR ordering.
   }
    \label{fig:Nmin_vs_aq}
\end{figure}

In Fig.~\ref{fig:Nmin_vs_aq}, we show the dependence of $N_{\text{min}}$ on the spin (upper panel) and mass ratio (lower panel) of the progenitors, for a system with primary mass $m_1=5 \times 10^6 M_{\odot}$ and luminosity distance $d_L=10\,\text{Gpc}$. The angles are fixed to $\theta=\psi=\iota=\pi/3$, $\phi=0$, and we consider equal spins. The discontinuous behavior that can be observed in both panels reflects the evolution of the modes ordering. In particular, the sudden jumps correspond to the exchange of two or more modes in the SNR ranking. 
We do not observe a clear trend of $N_{\text{min}}$ in any of the two cases. Overall, the impact of spin and mass ratio is mainly on the ordering of the modes, and only changes $N_{\text{min}}$ of $\pm 1$ mode, a smaller fluctuation with respect to the variability introduced by the orientation angles and shown in Fig.~\ref{fig:anlge_dep}. For very loud sources requiring a large \( N_{\text{min}} \), variations due to spin and mass ratio become larger but remain mostly subdominant compared to those from sky location and inclination.

\section{A mode-dependent exponential tapering in the frequency domain}\label{app:tapering}
\label{appD}
\begin{figure}[b]
          \centering
\includegraphics[width =0.45\textwidth]{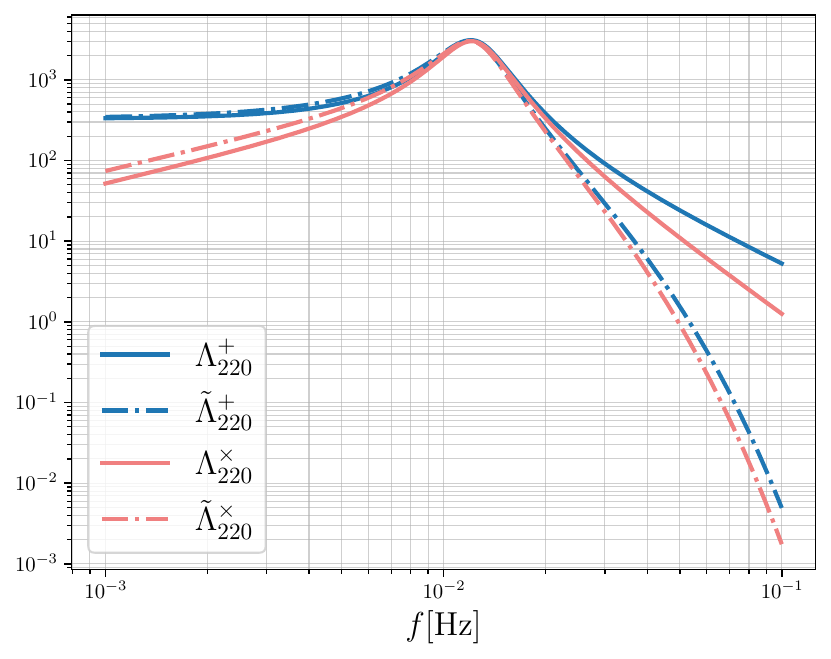}
 \caption{Comparison between the functions $\Lambda^+_{\ell m n}$ (continuous lines) and $\tilde{\Lambda}^+_{\ell m n}$ (dot-dashed lines) for the $220$ mode. The exponentially tapered functions recover the behavior of the standard combinations of Lorentzian combinations, while introducing a sharper cutoff at higher frequencies.}
\label{fig:TaperedLorentzian}
\end{figure}

\begin{figure}[t]
          \centering
\includegraphics[width =0.45\textwidth]{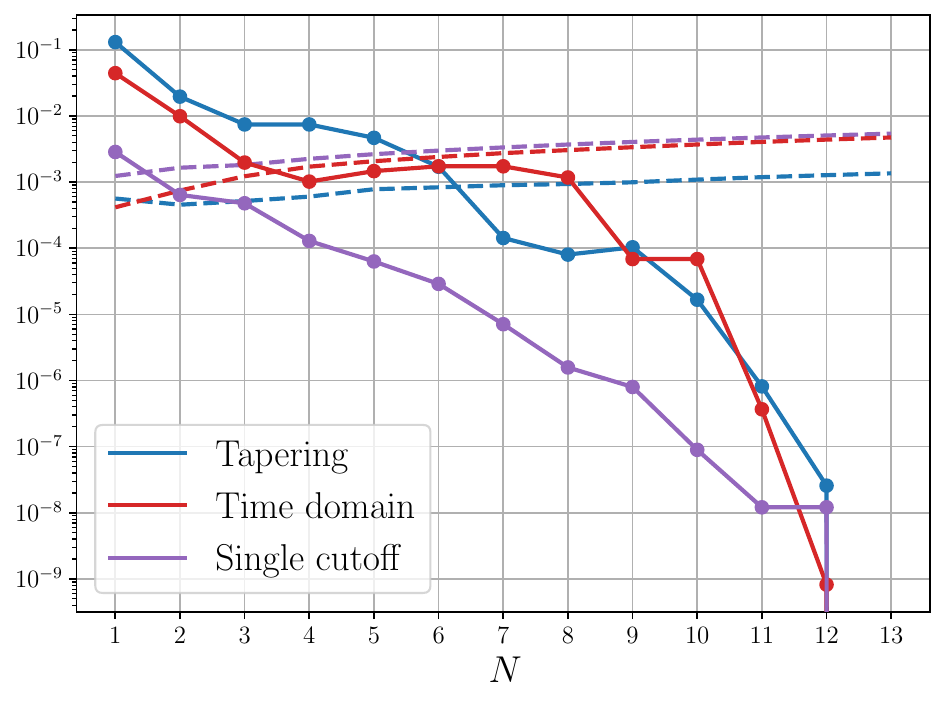}
 \caption{Mismatch obtained with three different approaches as a function of the number of modes. The remnant mass and spin of the system are respectively $2.66\times 10^7\,M_{\odot}$ and $0.60$, while the angles have been fixed to $\theta =3\pi/4$, $\phi = \psi = 0$, $\iota=\pi/3$ and $\varphi = \pi$. The dashed lines correspond to the threshold given by the right-hand side of Eq.~\eqref{eq:mismatch_criterion}. The blue, red and purple curves show the mismatch obtained respectively with the phenomenological tapering of Eq.~\eqref{eq:tapering}, with the time-domain approach described in Sec.~\ref{sec2}, and with the single high-frequency cutoff from PhenomA. Remarkably, the phenomenological tapering reproduces quite well the prediction from the time-domain analysis. On the other hand, both these choices depend on the window, while the PhenomA cutoff provides a conservative lower bound for $N_{\rm min}$. }
\label{fig:Mismatch_Comparison}
\end{figure}
In Sec. \ref{sec3} we discussed and motivated our choice for the high-frequency cutoff. In this appendix, we briefly present a different possible approach. In order to better mitigate the spectral leakage due to mirroring, we modify the frequency-domain templates with the replacement $\Lambda_{\ell m n}^{+,\times}\rightarrow \tilde{\Lambda}_{\ell m n}^{+,\times}$, introducing the functions
\begin{equation}
    \begin{split}
        &\tilde{\Lambda}_{\ell m n}^{+}=\frac{2L^+_{\ell m n }}{\exp{\left(\frac{2\pi f}{\omega_{\ell m n}}-1\right)}+1}+\frac{2L^-_{\ell m n }}{\exp{\left(\frac{2\pi f}{\omega_{\ell m n}}+1\right)}+1}\\
        &\tilde{\Lambda}_{\ell m n}^{\times}=\frac{2L^+_{\ell m n }}{\exp{\left(\frac{2\pi f}{\omega_{\ell m n}}-1\right)}+1}-\frac{2L^-_{\ell m n }}{\exp{\left(\frac{2\pi f}{\omega_{\ell m n}}-1\right)}+1}\,.
    \end{split}
    \label{eq:tapering}
\end{equation}
Recall that for $f\gg \omega_{\ell m n}, f\gg 1/\tau_{lmn}$, the high frequency tails due to mirroring scale as 
\begin{align}
 \Lambda_{\ell m n }^{+}\sim \tau_{\ell m n }^{-1}f^{-2} \,\, , \,\, \Lambda_{\ell m n }^\times\sim \omega_{\ell m n} \tau_{\ell m n }^{-1}f^{-3}.   
\end{align}
On the other hand, the phenomenologically tapered functions scale as 
\begin{align}
   \tilde{\Lambda}_{\ell m n }^+ \sim \dfrac{e^{-2\pi f/\omega_{\ell m n}}}{\tau_{\ell m n }f^{2}} \,\, , \,\,  \tilde{\Lambda}_{\ell m n }^\times \sim \dfrac{ \omega_{\ell m n}e^{-2\pi f/\omega_{\ell m n}}}{ \tau_{\ell m n }f^{3}} .
\end{align}
Thus, our phenomenological tapering mimics a faster than exponential fall-off (due to $1/f^2$ and $1/f^3$ for the $+$ and $\times$ polarizations respectively).
This approach is conceptually similar to the frequency-domain tapering of IMR waveforms~\cite{Husa:2015iqa, Khan:2015jqa,Mehta:2022pcn}. 
The comparison of these exponentially tapered functions and the standard combinations of Lorentzian functions is shown in Fig.~\ref{fig:TaperedLorentzian} for the $220$ mode and a BH mass of $10^6\,M_{\odot}$.

Our tapering can be motivated by constructing a time-domain Planck window~\cite{McKechan:2010kp} for each mode.
With a Planck window $W_{\rm Planck}(t)$, the high frequency fall-off is at most exponential, which can be sketched out in the following way.
The Fourier transform of a $C^{\infty}$ window function $W(t)$ with compact support, upon integration by parts $p$ times, can be expressed as $W(f) =W^{(p)}(f)/(i 2\pi f)^p $, where $W^{(p)}(f)$ is the Fourier transform of the $p$-th derivative of $W(t)$.
Since every derivative exists due to $W(t)$ being $C^{\infty}$ and $W(t)$ has compact support, we have that $|W(f)| \leq A/f^p$ (with $A$ being an overall integration factor), suggesting that $W(f)$ decays faster than every polynomial in frequency.
In other words, $|W(f)|$ can at most have an exponential fall-off. This implies that the $C^{\infty}$ Planck window will satisfy $|W_{\rm Planck}(f)| \leq A \exp(- B f)$, where $B \sim \Delta t$, with $\Delta t$ being the characteristic transition time of the window.
With our phenomenological tapering, for each mode, the corresponding mode-dependent Planck window would then have an exponential fall-off $B_{\ell m n} \sim 1/\omega_{\ell m n}$. 
However, by tuning the mode-dependent transition time $\Delta t_{\ell m n}$, in practice, one can get a faster exponential fall-off with a mode-dependent Planck window.

As a corollary, we can also obtain the frequency fall-off when $W(t)$ is discontinuous in its $p$-th derivative at $t=t_0$.
The Fourier transform $W^{(p)}(f)$, without loss of generality, will contain a term 
\begin{align}
    W^{(p)}(f) \supset \int_{t_0}^{\infty} g(t) \exp^{-i 2\pi f t},
\end{align}
where $g(t)$ is $C^{\infty}$ and has compact support.
Upon integrating by parts, we have that 
\begin{align}
    W^{(p)}(f) \supset g(t_0) \dfrac{e^{-i 2\pi f t_0}}{i 2\pi f} \left[ 1+\mathcal{O}(f^{-1}) \right],
\end{align}
which is also equivalent to Taylor expanding $g(t)$.
Given that $W(f) =W^{(p)}(f)/(i 2\pi f)^p $, to leading order for the frequency fall-off, we simply have that $|W(f)| \sim 1/f^{p+1}$.
For the commonly used Heaviside window\footnote{Note that although the Heaviside window doesn't vanish for $t \rightarrow \infty$, the result still applies because $h(t) W(t)$ is compact throughout the domain, with $h(t)$ being the GW strain.} for ringdown analysis, this means that the spectral leakage fall-off is only $1/f$.
Instead, we use a frequency domain ringdown model obtained from mirroring~\cite{Flanagan:1997kp,Berti:2005ys} at $t=t_0$. 
Although mirroring is not identical to a window function multiplying the damped sinusoid, due to the mirrored waveform being $C^0$ at $t=t_0$, it mimics the use of a $C^0$ window, and thus leads to a $1/f^2$ fall-off (cf. Eq.~\eqref{eq:ringdownModelFreqDomain}).
Note, however, that $\Lambda^{\times}_{\ell m n}$ scales as $1/f^3$ instead of $1/f^2$ as is the case for $\Lambda^{+}_{\ell m n}$. This is due to the fact that $\Lambda^{\times}_{\ell m n}$ is an anti-symmetric combination which makes the leading $1/f^2$ term cancel out, leaving the next-to-leading $1/f^3$ scaling. 

\begin{figure}[b]
          \centering
\includegraphics[width =0.45\textwidth]{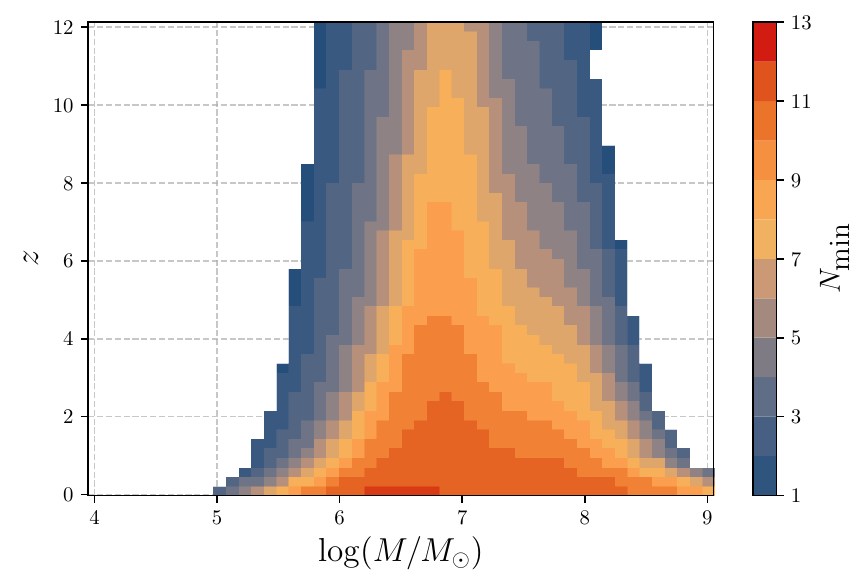}
 \caption{ $N_{\rm min} $ obtained with the criterion of Eq.~\eqref{eq:sys_stat_accuracy} and using the tapered functions in \eqref{eq:tapering} as a function of primary mass and redshift. While the qualitative behavior is similar to the one in Fig.~\ref{fig:Waterfall}, it is evident how $N_{\rm min}$ is boosted to higher values with this approach.}
\label{fig:Waterfall_Tapering}
\end{figure}

A major caveat of using the phenomenological tapering is that one cannot simply construct a time-domain window $W^{(\ell mn)}(t)$ for each mode. 
This is because \( W^{(\ell m n)}(t) \) would also convolve with modes \( (\ell', m', n') \neq (\ell, m, n) \), which are not included in our phenomenologically tapered frequency-domain model.
However, we checked that results obtained with our phenomenologically tapered frequency-domain model agree well with a pure time-domain analysis.

To enable the comparison, we generate the waveform in the time domain as described by Eq.~\eqref{eq:QNMModelTimeDomain}, including the low-frequency approximation for the LISA response. In order to compute the inner product of two strains in the time domain, one needs to compute the two-point correlation function $C(t'-t)$ from the frequency-domain noise PSD, and combine it with the two strain series as described, for example, in Ref.~\cite{Pitte_24}.
%

For a source with remnant mass and spin of $2.66\times 10^7\,M_{\odot}$ and $0.60$ respectively, and angles $\theta =3\pi/4$, $\phi = \psi = 0$, $\iota=\pi/3$ and $\varphi = \pi$, we apply the mismatch criterion given by Eq.~\eqref{eq:mismatch_criterion} with the two methods (time and frequency domain). We show the result in Fig.~\ref{fig:Mismatch_Comparison}. The mismatch $1-{\cal M}$ is shown by solid lines,  while its threshold (right-hand side of Eq.~\eqref{eq:mismatch_criterion}) is shown by dashed lines. The comparison is presented for different models, namely: the time domain (TD) mismatch in red; the frequency domain mismatch with a common single cutoff in purple; and the frequency domain mismatch with optimal tapering in blue. With the cutoff mismatch, two modes are sufficient for a  correct estimation of the parameters, while seven modes are needed with the optimal taper. 
For the time-domain mismatch, four modes are sufficient to fulfill this requirement. We can therefore conclude that with the single cutoff, we can conservatively estimate the minimum number of modes needed for unbiased parameter estimation with damped-sinusoids waveforms.


\bibliographystyle{apsrev4-1}

\bibliography{Bibliography}

\end{document}